\documentclass[aps,twocolumn,amsmath,amssymb,preprintnumbers]{revtex4}
\usepackage{amsmath} \usepackage{amsfonts} \usepackage{amssymb}
\usepackage{bbm}
\usepackage{epsfig}
\usepackage{graphics}
\usepackage{graphicx}
\newcommand{\be}{\begin{equation}}
\newcommand{\ee}{\end{equation}}
\newcommand{\ba}{\begin{eqnarray}}
\newcommand{\ea}{\end{eqnarray}}
\newcommand{\nn}{\nonumber}

\newcommand{\bv}{\bar{\varphi}}

\newcommand{\mn}{_{\mu\nu}}
\newcommand{\hmn}{_{\hat\mu\hat\nu}}
\newcommand{\4}{^{(4)}}

\begin{document}

\title[ ]{Naturalness of exponential cosmon potentials and the cosmological constant problem}

\author{C. Wetterich}
\affiliation{Institut  f\"ur Theoretische Physik\\
Universit\"at Heidelberg\\
Philosophenweg 16, D-69120 Heidelberg}

\begin{abstract}
We discuss the naturalness of exponential potentials for quintessence, showing that the resulting almost flat direction in the space of scalar fields, as well as the small time dependent cosmon mass, can be related to an anomalous dilatation symmetry. We argue that the physics of the cosmological constant is Planck scale physics, and comment on the role of quantum fluctuations. We present three higher dimensional scenarios how a runaway of the ``cosmological constant'' to zero can be combined with stable particle physics properties, leading at most to small variations of the ``fundamental couplings''. 
\end{abstract}

\maketitle

\section{Introduction}
For scalar fields with an effective potential relaxing exponentially to zero for large field values, the presence of a homogeneous dark energy component in the universe has been predicted \cite{CWQ}. The dynamics of the cosmological scalar field - the cosmon - leads to cosmological scaling solutions \cite{CWQ}, \cite{SC} where dark energy decreases with time at the same pace as matter or radiation. This predicts for the present cosmological epoch dark energy and matter of a comparable order of magnitude. 
If the scaling can be stopped by some cosmological trigger event, as neutrinos with growing mass becoming non-relativistic at low redshift\cite{ABW}, a cosmology that is compatible with present observation can result.

The naturalness of an exponential cosmon potential in the presence of quantum fluctuations has been questioned on several grounds. The first criticism is that quantum fluctuations suggest that the potential should relax to some constant rather than to zero. This is the ``cosmological constant problem''. The second question concerns the naturalness of the exponential shape. A naive quantum calculation suggests that the exponential shape is stable as far as the quantum fluctuations of the cosmon are concerned, but becomes unnatural in the presence of a cosmon coupling to matter \cite{DJ}. Finally, it has been challenged that the small cosmon mass may be unnatural \cite{CM}. Indeed, for the scaling solution the cosmon mass decreases with time and is of the order of the Hubble parameter \cite{CWAA}, much smaller than other known particle masses. 

We will argue in this paper that all these criticisms are not justified, even in presence of a coupling of the cosmon to dark matter, neutrinos or atoms. The basic reason is that a too naive computation of the quantum fluctuations does not respect an important underlying symmetry, namely dilatation symmetry. In fact, the cosmon is the pseudo-Goldstone-boson of a spontaneously broken dilatation symmetry. In absence of a dilatation anomaly the cosmon potential would be flat and its mass would vanish. Furthermore, dilatation symmetry has the special property that it gets restored if a fixed point is approached. In this case the anomaly vanishes asymptotically as a result of the dynamics. This is precisely what happens for the cosmological scaling solution for which the cosmon mass goes asymptotically to zero.

In the context of a unified theory one expects the presence of many scalar fields $\bar\varphi_i$. For example, dimensional reduction of a higher dimensional theory will lead to infinitely many scalar fields. These scalars may be singlets with respect to the $SU(3)\times SU(2)\times U(1)$ gauge symmetry of the standard model of particle physics or not. The dynamics of these scalar fields will be determined by an effective potential $U(\bar\varphi_i)$, which obtains by including all effects from quantum fluctuations. An anomalous dilatation symmetry will lead to a ``valley'' or ``almost flat direction'' in the effective potential, denoted by fixed functions $\bar\varphi_i(\varphi)$. Here the ``coordinate'' along the valley, $\varphi$, will be associated with the cosmon field. We will present rather general arguments why the effective cosmon potential takes an exponential form
\ba\label{1}
\bar V(\varphi)&=&U\big(\bar\varphi_i(\varphi)\big)=\lambda_* M^4
+V(\varphi),\nn\\
V(\varphi)&=&M^4\exp\left(-\alpha\frac{\varphi}{M}\right).
\ea
Here $M$ is the reduced Planck mass and $\alpha$ a dimensionless parameter. The dimensionless constant $\lambda_*$ determines the size of the cosmological constant. Realistic cosmology requires $\lambda_*\lesssim 10^{-120}$. 

We also argue in favor of $\lambda_*=0$. The asymptotic vanishing of the cosmon potential can not be explained as a result of an anomalous dilatation symmetry. It is rather connected to the stability properties of the theory and we present three scenarios of higher dimensional cosmology that realize $\lambda_*=0$.

In the first part of this paper (sects. \ref{dilatationsymmetry}-\ref{dilatationsymmetryhigher}), we answer fifteen general questions concerning the possibility of a natural solution to the cosmological constant problem and the role of quantum fluctuations. This includes the possibility of a cosmon coupling to baryons, neutrinos or dark matter particles, and therefore the issue of time dependence of the particle physics couplings like $\alpha$ or $m_e/m_p$. 

The second part of this paper (sects. \ref{simplemmodel}-\ref{adjustinginternal}) addresses the cosmological constant problem in a higher dimensional context. Realistic cosmologies need not only a dynamical dark energy that approaches zero asymptotically. Also the couplings must become almost time independent in order not to contradict tight observational bounds. While each aspect separately is easily realized, the challenge resides in the simultaneous realization of both properties. We present three scenarios that could give rise to realistic cosmologies in this sense. The {\em anomalous runaway} (sect. \ref{dilatationinhigher}) is based on a higher dimensional theory that has only dimensionless couplings. Their scale dependence is given by an anomalous dimension, thus leading to a dilatation anomaly. The {\em geometrical runaway} for warped branes (sect. \ref{geometricalrunaway}) is characterized by an increasing size of the internal dimensions in units of the Planck length, while the scale governing the particle physics is related to the geometry of a brane and becomes time independent. For the {\em adjustment of internal geometry} (sect. \ref{adjustinginternal}), the length scale characteristic for internal curvature increases to infinity, while the characteristic size of the internal space approaches a constant. All three scenarios lead, after dimensional reduction and in the Einstein frame, to an exponential cosmon potential.

\section{Dilatation symmetry and pseudo Goldstone boson}
\label{dilatationsymmetry}

\bigskip\noindent
{\em (1) Why is there an almost flat direction?}

For a realistic cosmology the present value of $\varphi$ obeys $\alpha\varphi(t_0)/M\approx 276$. Therefore the derivative of $V$ is much smaller than naively expected from the characteristic scales for gravity, $M\approx 10^{18}$ GeV, weak interactions, $d\approx 175$ GeV, or strong interactions, $\Lambda_{QCD}\approx 200$ MeV, i.e.
\ba\label{2}
-M\frac{\partial V}{\partial \varphi}=&&\alpha V\approx \alpha\rho_h\approx\alpha(2\cdot 10^{-3} eV)^4\nn\\
&&\ll M^4,~d^4,~\Lambda^4_{QCD}.
\ea
The reason for the existence of a flat direction is the spontaneous breaking of the global dilatation symmetry. In the limit of a vanishing dilatation anomaly an exactly flat direction would exist, $V'=\partial V/\partial \varphi =0$. The cosmon becomes the Goldstone boson that is always generated by a spontaneously broken global symmetry - it plays the role of the dilaton. The dilatation transformation induces a shift of the dimensionless field $\varphi/M\to \varphi/M+\vartheta$. There is only one difference to a spontaneously broken compact $U(1)$ symmetry, namely that $\vartheta$ is not a periodic variable. Dilatations correspond to a non-compact symmetry group such that $\vartheta$ can take arbitrary values. The flat direction is not a closed line in field space.

\bigskip\noindent
{\em (2) Why do quantum fluctuations of QCD degrees of freedom not lift the flatness and induce a contribution $MV'\sim\Lambda^4_{QCD}$?}

The characteristic scale for QCD is not the characteristic scale for the explicit breaking of the dilatation symmetry - the dilatation anomaly. In this context one has to distinguish between the ``fundamental dilatation symmetry'' and an effective low energy dilatation symmetry in absence of gravity. For energy scales sufficiently below the unification scale, particle physics can be described by the degrees of freedom of the standard model. The effective action for these ``low energy degrees of freedom'' exhibits an approximate effective dilatation symmetry. The latter is only broken by the running of the dimensionless couplings, as the strong gauge coupling $g_s$, and by the mass term in the effective potential for the Higgs scalar. (Technical naturalness of a small Fermi scale as compared to the unification scale can be associated with this approximate effective dilatation symmetry \cite{CWGH}). Due to the running strong coupling there is indeed a contribution to the ``anomaly'' of the effective dilatation symmetry of the order $\Lambda^4_{QCD}$. 

However, the issue is different for the fundamental dilatation symmetry with which we are concerned here. In a unified theory the scale $\Lambda_{QCD}$ obeys $\Lambda_{QCD}=M_x\exp\big(-c/g^2_s(M_x)\big)$, with $M_x$ the unification scale and $g_s(M_x)$ the running strong coupling at this scale. In models with a fundamental dilatation symmetry both the Planck mass $M$ and the unification scale $M_x$ are (approximately) proportional to a field $\chi$ that scales under dilatation transformations. If the proportionality is exact, and $g^2_s(M_x)$ is independent of $\chi$, the fundamental dilatation symmetry remains unbroken. The running strong coupling induces in this case no anomaly of the fundamental dilatation symmetry, since also $\Lambda_{QCD}$ is proportional to $\chi$. In the limit of a vanishing dilatation anomaly the effective potential $U(\bar\varphi_i)$ will therefore exhibit a flat direction, even in presence of the QCD-fluctuations. These fluctuations will, of course, contribute to $U$. However, they will only modify the precise location of the flat direction in field space, i.e. the functions $\bar\varphi_i(\varphi)$. The flatness of the Goldstone direction remains untouched. This argument generalizes, of course, to be electroweak fluctuations.

We can phrase these statements differently: In the limit of an exact fundamental dilatation symmetry, the QCD-contribution to the fundamental dilatation anomaly must be precisely cancelled by the contributions from other fluctuations. This cancellation involves no fine tuning of parameters since it is enforced by symmetry. This feature is often overlooked when one tries to estimate the issue of quantum contributions to $V'$ or $V''$ from an effective theory covering only a certain range in momentum. We will demonstrate the shortcomings of ''effective theory computations'' in a simple and well understood model. 

Indeed, in this respect the situation is completely analogous to the Goldstone boson arising from the spontaneous breaking of a compact $U(1)$-symmetry. Consider two complex fields $\bar\varphi_1$ and $\bar\varphi_2$ transforming as $\bar\varphi_i\to \bar\varphi_ie^{i\theta}$. A polynomial expansion of a $U(1)$ invariant classical or effective potential has the general form
\be\label{3}
U=\mu^2_{ij}\bar\varphi^*_i\bar\varphi_j+\frac12\lambda_{ijkl}
\bar\varphi^*_i\bar\varphi_j\bar\varphi^*_k\bar\varphi_l,
\ee
with $\mu^2_{ji}=\mu^{2*}_{ij},~\lambda_{jilk}=\lambda^*_{ijkl},~\lambda_{ijkl}=\lambda_{klij}$. Without loss of generality the solution of the field equation (for homogeneous static $\bar\varphi)$
\be\label{4}
\mu^2_{ij}\bar\varphi_{j,0}+\lambda_{ijkl}\bar\varphi_{j,0}\bar\varphi^*_{k,0}\bar\varphi_{l,0}=0,
\ee
can be written in the form
\be\label{5}
\bar\varphi_{1,0}=a,~\bar\varphi_{2,0}=b e^{i\gamma},
\ee
with $a,b$ real and positive and $\gamma$ an angle. The deviations from the potential minimum, $\delta\bar\varphi_i=\bar\varphi_i-\bar\varphi_{i,0}$, contain a flat direction labeled by $\vartheta$
\be\label{6}
\delta\bar\varphi_{i,G}=\bar\varphi_{i,0}e^{i\vartheta}-\bar\varphi_{i,0}.
\ee
Here $\vartheta$ plays the role analogous to the dilaton $\varphi$, i.e. the cosmon in absence of a dilatation anomaly.

Consider now for a classical potential $U$ the case $a\gg b$. In leading order the Goldstone boson corresponds to the imaginary part of $\bar\varphi_1,~g=Im\delta\bv_1=a\vartheta+O(\vartheta^3)$. Let us take the limit $b\to 0$ (this can actually always be achieved in our example by appropriate field conventions) and expand up to quadratic order in $\delta\bv_2=\bv_2$
\ba\label{7}
U&=&\left[\mu^2_{22}+(a^2+g^2)\big(\lambda_{1122}+\frac12(\lambda_{1221}+\lambda^*_{1221})\big)\right]
\bv^*_2\bv_2\nn\\
&&+\frac12\big[(a^2-2ia g-g^2)\lambda_{1212}\bv^2_2+c.c.\big]\nn\\
&&-\big[g^2(a-ig)\lambda_{1112}\bv_2+c.c.\big].
\ea
Eq. \eqref{7} contains a mass matrix for the two real fields which correspond to the complex field $\bv_2$, whereas $g$ is the massless Goldstone boson. In addition, the linear description $\bv_1=a+ig$ generates cubic and quartic couplings involving $g$. (In a non-linear field basis the Goldstone boson has only derivative couplings.) Let us further consider parameters where one of the mass eigenvalues for $\bv_2$ (with eigenvector denoted by $\sigma$) is much smaller than the other and also much smaller than the mass for the real part of $\delta\bv_1$. For momenta below an ``ultraviolet cutoff'' $\Lambda^2\sim\lambda a^2$ one may consider an ``effective low energy theory'' for $\sigma$ and $g$. In our analogue $\sigma$ corresponds to the QCD-degrees of freedom and $g$ to the cosmon.

One may be tempted to compute the effective potential and the characteristic size of the mass term for $g$ by integrating out the fluctuations of $\sigma$ in a momentum range $q^2<\Lambda^2$. Due to the cubic and quartic couplings between $\sigma$ and $g$ such a calculation would yield a mass term $\sim\lambda\Lambda^2 g^2$ or $\lambda^2a^2g^2$. Such a result is, of course, grossly erroneous since the quantum fluctuations of the effective theory do not respect the $U(1)$ symmetry. A full computation of the quantum effects respects the $U(1)$ symmetry. It only changes the values of $\mu^2$ and $\lambda$ from their classical values to the quantum values (and introduces further $U(1)$-invariant ``higher order terms'' in $U$). As a consequence, the values of $a,b,\gamma$ are shifted from their classical values to quantum values. This only affects the location of the flat direction, while the Goldstone boson remains exactly massless. One concludes that the contribution to the Goldstone boson mass from fluctuations of the effective theory is exactly cancelled by other contributions of quantum fluctuations of the full theory. 

We may take this simple example as a lesson: in presence of a spontaneously broken symmetry it is not legitimate to consider the contribution of quantum fluctuations of an effective low energy theory as representing the characteristic size of the quantum effects for the full theory.

\bigskip
\noindent
{\em (3)\quad How can the cosmon mass be many orders of magnitude smaller than $\Lambda_{QCD}$ or $\Lambda^2_{QCD}/M$?}

The answer to this question is essentially the same as above - the small mass is a consequence of a spontaneously broken global symmetry with a small anomaly. The present cosmon mass is easily computed by taking the second derivative of eq. \eqref{1}
\be\label{8}
m^2_\phi=\alpha^2 V/M^2\approx 3\alpha^2\Omega_{h,0}H^2_0
\ee
with $\Omega_{h,0}\approx 3/4$ the present dark energy fraction and $H_0$ the present value of the Hubble parameter. We can associate the reduced Planck mass $M$ with the scale of spontaneous breaking of dilatation symmetry. The potential $V$ plays the role of the dilatation anomaly. In presence of an anomaly the cosmon becomes a pseudo-Goldstone-boson and its mass involves the typical ratio between the anomaly and the scale of spontaneous breaking, $m_\phi\sim\sqrt{V}/M$. This is completely analogous to the axion with anomaly $\Lambda^4_{QCD}$ and decay constant $f,~m_a\sim\Lambda^2_{QCD}/f$. The only difference is that the anomaly $V$ is not related to $\Lambda_{QCD}$ and actually changes with increasing $\varphi$. For $\varphi\to\infty$ the anomaly vanishes and the cosmon becomes a massless Goldstone boson.

It has been argued that a cosmon coupling to fermions, for example to neutrinos, would generate a mass much larger than $H_0$. A coupling $\sim\beta(\varphi/M)m_\nu\bar\psi\psi$ induces a $\varphi$-dependent neutrino mass and a corresponding cubic coupling between the cosmon and neutrinos. A naive computation of quantum fluctuations of neutrinos with momenta smaller than $\Lambda\approx M$ would yield a contribution to the cosmon mass $\Delta m^2_\phi\sim(\beta m_\nu/M)^2\Lambda^2\approx\beta^2 m^2_\nu$, which is many orders of magnitude larger than $H^2_0$ unless $\beta$ is tiny. The shortcoming in this argument is precisely the same as in the above discussion of the global $U(1)$ symmetry with the fields $\bv_1,\bv_2$: the computation within the effective low energy theory does not respect the symmetry which is responsible for the small cosmon mass.

\bigskip\noindent
{\em (4)\quad Why does the dilatation anomaly vanish for $\varphi\to\infty$?} 

The vanishing of the dilatation anomaly $V(\varphi\to\infty)=0$ is linked to a fixed point behavior in the renormalization flow. In a theory without explicit mass scales the dilatation anomaly arises from the scale dependence of running dimensionless couplings. Assume that all dimensionless couplings take constant values at fixed points. Then the dilatation anomaly must vanish, since it is related to the running of the couplings. As the couplings approach fixed points, the dilatation symmetry is restored. In this case the anomaly is ``switched off'' dynamically \cite{Veneziano}. While our discussion of the first three questions is general for all spontaneously broken continuous global symmetries, the ``switching off'' of the anomaly is particular to dilatation symmetry.

This issue is most easily understood in a field basis where dilatation transformations are realized linearly and the dilatation symmetry is manifest in the action. In this formulation all parameters with dimension of mass are given by $h\chi$, with $h$ a dimensionless coupling and $\chi$ a real scalar field. For example, a dilatation symmetric model for gravity $(g_{\mu\nu})$, a scalar $(\chi)$, and a fermion $(\psi)$ has a Lagrange density
\ba\label{9}
-{\cal L}&&=\sqrt{g}\left\{\frac{Z_\chi-6}{2}\partial^\mu\chi\partial_\mu\chi+
\lambda\chi^4
-\frac12\chi^2R\right.\\
&&\left.+i(\bar\psi_L\gamma^\mu D_\mu\psi_L+\bar\psi_R\gamma^\mu D_\mu\psi_R)+
(h\chi\bar\psi_R\psi_L+h.c.)\right\}.\nn
\ea
Here we have chosen a normalization of the scalar field such  that  $\chi$ corresponds to the reduced Planck mass. We note that for a time varying $\chi$ the Planck mass $\tilde M=\chi$ changes with time. However, the effective fermion mass $m_\psi=h\chi$ also changes. Only dimensionless ratios are observable. For $\chi$-independent $h$ the ratio $m_\psi/\tilde M=h$ remains constant. The ``dilatation symmetric frame'' \eqref{9} is called the Jordan frame. It is related by a Weyl transformation of the metric to the Einstein frame where $M$ is a constant. The field $\varphi$ corresponds to $\ln\chi$. A multiplicative scaling of $\chi$ corresponds to a shift in $\varphi$. Therefore the dilatation transformations act linearly on $\chi$ (an infinitesimal transformation is $\delta\chi=\epsilon\chi)$, whereas a formulation with $\varphi$ corresponds to the ``nonlinear $\sigma$-model'' for Goldstone bosons.

In a quantum theory we should use an effective Lagrangian. Due to the effects of quantum fluctuations the dimensionless couplings $\lambda,Z_\chi,h$ become running couplings, depending on $\chi/\mu$, with $\mu$ the renormalization scale. In the limit of $\chi$-independent couplings the effective Lagrangian becomes dilatation symmetric. In consequence, the dilatation anomaly involves the $\beta$-functions of the running couplings, i.e.
\be\label{10}
\frac{\partial\lambda}{\partial\ln\chi}=\beta_\lambda.
\ee
Performing the Weyl transformation and using $\varphi/M\sim\ln(\chi/M)$, we indeed find a flat direction in the effective potential for $\varphi$ if $\beta_\lambda$ vanishes.

In cosmology, there are two fundamentally different possible behaviors for $\chi$. Either $\chi$ settles for large time at some constant value, or it evolves towards infinity. We will be interested in the second class, the ``runaway solutions''. Depending on the form of $\beta_\lambda$, the evolution of the coupling $\lambda$ for the runaway solution has two alternatives. Either $\lambda$ approaches a fixed point $\lambda_*$ where $\beta_\lambda$ vanishes, $\beta_\lambda(\lambda_*)=0$. Or $\lambda$ runs to infinity. We assume here the existence of a fixed point that is approached for $\chi\to \infty$. This implies that the dilatation anomaly vanishes for $\varphi\to\infty$, and we thus have demonstrated the generic behavior $V(\varphi\to\infty)=0$.

\bigskip\noindent
{\em (5)\quad  Why does $V(\varphi)$ vanish exponentially for $\varphi\to\infty$?}

The exponential behavior is a result of an anomalous dimension characterizing the approach to a fixed point. Typically, the $\beta$-function is an analytic function of $\lambda$ in the vicinity of $\lambda_*$. The lowest term in a Taylor expansion is given by the ``anomalous dimension'' $A$,
\be\label{11}
\beta_\lambda=-A(\lambda-\lambda_*).
\ee
Eq. \eqref{10} implies then
\be\label{12}
\lambda(\chi)=\lambda_*+(\chi/\mu)^{-A}.
\ee
The effective Lagrangian for $\chi$ reads in the Einstein frame
\ba\label{13}
-{\cal L}=\frac12 Z_\chi\partial^\mu\chi\partial_\mu\chi\frac{M^2}{\chi^2}+
\left(\lambda_*+\left(\frac\chi\mu\right)^{-A}\right)\chi^4\frac{M^4}{\chi^4}\nn\\
=\frac12\frac{Z_\chi}{Z_0}\partial^\mu\varphi\partial_\mu\varphi+M^4
\left[\lambda_*+\exp\left(\exp(-\alpha\frac\varphi M\right)\right].
\ea
The kinetic term for the field
\be\label{14}
\varphi=M\sqrt{Z_0}\ln(\chi/\mu)
\ee
is canonical for a constant value $Z_\chi=Z_0$ and $\alpha$ is related to $A$ by
\be\label{15}
\alpha=AZ^{-1/2}_0.
\ee
For a $\chi$-dependent $Z_\chi$ one may either keep the exponential form of the potential and use a nonstandard kinetic term with ``kinetial'' $k^2(\varphi)=Z_\chi/Z_0$, or one may rescale $\varphi$ in order to achieve a standard kinetic term, with corresponding modifications of the potential. For $Z_\chi$ only slowly varying, for example close to a fixed point, the effective potential is then approximately of exponential shape.

A naive computation of the quantum fluctuations from fermions in an effective low energy theory (for example realized by small $h$) would suggest deviations from the exponential form of the cosmon potential. Such a computation is misleading, however. Typically, both $\lambda$ and $h$ may approach fixed points as $\chi\to\infty$ for a runaway solution. The behavior for $\chi\to\infty$ is then governed by a stability matrix $S$
\be\label{16}
\frac{\partial}{\partial \ln\chi}\left(\begin{array}{l}\lambda-\lambda_*\\h^2-h^2_*\end{array}\right)=-
S\left(\begin{array}{l}\lambda-\lambda_*\\h^2-h^2_*\end{array}\right).
\ee
If the fixed point is approached for $\chi\to\infty$, all eigenvalues of $S$ are positive. The dominant deviation from the fixed point for large $\chi$ is given by the lowest eigenvalue. If the corresponding eigenvector contains nonvanishing components of both $\lambda-\lambda_*$ and $h^2-h^2_*$, both couplings approach their fixed points with the same power law \eqref{12}. On the other hand, for block diagonal $S$ the respective powers correspond to the two eigenvalues of $S$ and may be different. In both cases an  exponential form of the cosmon potential follows.

At this point we have answered the key issue to which we refer in the title. A dilatation symmetric theory with an anomaly due to running dimensionless couplings leads naturally to an exponential form of the cosmon potential, if fixed points are approached for a cosmological runaway solution. The quantum fluctuations are responsible for the nontrivial $\beta$-function and determine the properties of the stability matrix $S$. They generate the exponential potential, rather than destroying it. We note that on a classical level exponential potentials have been found as arising typically from higher dimensional \cite{SW} or supergravity theories. Here we have presented a deeper rooting on the quantum level.

\bigskip\noindent
{\em (6) \quad In which frame should  the quantum fluctuations be computed?}

The field equations derived from the quantum effective action are exact. We can therefore freely choose coordinates in field space that are convenient for their solution. The Jordan frame and the Einstein frame are completely equivalent on this level \cite{CWQ}. Physical observables can be formulated in a frame-independent way \cite{FI}. For a computation of the quantum fluctuations, however, a functional integral has to be performed. In principle, a change of field variables remains possible, but it often leads to a complicated Jacobian from the functional measure. A quantum calculation with a simple measure in a dilatation symmetric frame (Jordan frame) will induce a very complicated measure in the Einstein frame. If one uses, instead, a naive measure in the Einstein frame one will get a different result. For a computation related to the issue of dilatation symmetry and its anomaly, i.e. for a computation of the beta-functions for the running dimensionless couplings, it seems compulsory to perform it in the Jordan frame. In the Einstein frame the issue of running dimensionless couplings is not visible anymore and there is little chance to produce a sensible result if a naive measure is used.

\section{Time varying fundamental ``constants''}
\label{timevarying}

\bigskip\noindent
{\em (7) \quad Can we explain why ``fundamental constants'' change very little for a runaway solution?}

The ``fundamental constants'' like the electron or proton mass or the fine structure constant depend on the cosmological values of scalar fields. For example, a vanishing value of the Higgs scalar in early cosmology implies a similar strength for the electromagnetic and weak interactions. Before nucleosynthesis, however, most scalar fields are assumed to have settled to almost constant values, corresponding to a (partial) minimum of the effective potential. In the presence of a runaway solution, however, the cosmon field $\varphi$ continues to change even in recent cosmology. In principle, the fundamental couplings depend on $\varphi$. Due to the time variation of $\varphi$ one may expect a substantial variation of the couplings even in present cosmology. Explaining why this variation is small is one of the major challenges for realistic quintessence models \cite{CWQ1}. Indeed, the interaction between atoms or electrons mediated by the cosmon field must be substantially weaker than gravity, whereas many models investigated in the context of string theories or higher dimensional theories seem to predict a strength similar to gravity. 

Let us emphasize that only the variation of dimensionless couplings or dimensionless ratios of masses is observable. We may measure a variation of $m_e/M$ whereas an individual variation of the electron mass $m_e$ or the Planck mass $M$ is not observable. In our dilatation symmetric setting the time variation of mass ratios is directly related to the running of the dimensionless couplings, i.e. $m_\psi/M=h$. If all dimensionless couplings take fixed point values, the dilatation symmetry becomes exact. In this case, the couplings are constant. A time variation of the couplings is thus directly linked to the dilatation anomaly. If fixed points are approached asymptotically for $\varphi\to\infty$, the $\varphi$-dependence of the couplings becomes exponentially weak
\be\label{17}
h^2=h^2_*+\tilde c\left(\frac{\chi}{\mu}\right)^{-B}
=h^2_*+c\exp\left(-\beta\frac{\varphi}{M}\right)~,~\beta=BZ^{-1/2}_0.
\ee

For $\beta=\alpha$ or $\beta$ of a similar size as $\alpha$, the coupling $h$ would be very close to its fixed point value. With $\exp(-\alpha\varphi/M)\approx 10^{-120}$ for the present value of $\varphi$ in a realistic quintessence cosmology, a similar suppression factor would govern the deviation of $h$ from $h_*$. Runaway solutions with a fixed point for all couplings offer a simple scenario for all couplings to become essentially static in the recent cosmological epoch.

\bigskip\noindent
{\em (8) \quad Is there a chance to observe time varying couplings in present cosmology?}

A similar size of $\alpha$ and $\beta$ is, however, not the only possibility. For $\beta$ sufficiently small as compared to $\alpha$, the coupling $h$ is not yet extremely close to its fixed point. A time variation in an observable range becomes realistic in this case. The condition $\beta\ll\alpha$ requires that one of the eigenvalues $B$ of the stability matrix $S$ \eqref{16} is much smaller than $A$. In turn, this necessitates for $\lambda$ an evolution equation of the type of eq. \eqref{11}, where $A$  may depend on $h$, but no term $\sim h^2-h^2_*$ is present in a linear expansion of $\beta_\lambda$ around the fixed point. The eigenvalue $B$ is then irrelevant for the approach of $\lambda$ to its fixed point.

Let us pursue this scenario where some of the particle physics couplings show a slow evolution, whereas $\lambda$ approaches a fixed point much faster. In this case, the evolution of some particle physics couplings may not be governed by a fixed point - for example the relative change since the end of inflation could be of the order one only. Instabilities may be approached by the flow, like the vanishing of some mass, as in the scenario of ``growing neutrinos'' \cite{ABW}. We conclude that a small variation of fundamental couplings in an observable range seems perfectly possible, but it requires the decoupling of the comparatively fast flow of $\lambda$ from the slow flow of some particle physics couplings.

\section{Vanishing cosmological constant}
\label{vanishingcosmological}

\bigskip\noindent
{\em (9) \quad Why does dark energy vanish asymptotically?} 

In the remainder of this note we will present arguments in favor of a fixed point $\lambda_*=0$ that is approached for $\chi\to\infty$. The existence of such a fixed point, combined with an anomalous dimension dimension $A>0$, would solve the problem of the cosmological constant \cite{CWQ}. In this case the fourth power of the Planck mass $\chi^4$ increases faster than the effective potential $\tilde U(\chi)$.  In other words, it is a sufficient condition for a solution of the cosmological constant problem that $\tilde U(\chi)$ increases for $\chi\to\infty$ with a power smaller than four, $\tilde U(\chi)\sim\chi^{4-A}$. Then the ratio $\tilde U/\chi^4$ vanishes asymptotically according to eq. \eqref{1}. Runaway solutions lead to a vanishing dark energy for $\chi\to\infty$, unless stopped by some particular cosmological event. 

The zero value for the fixed point, $\lambda_*=0$, is not a consequence of dilatation symmetry. Runaway solutions highlight the importance of a fixed point where dilatation symmetry is dynamically restored. But dilatation symmetry does not tell the value of the fixed point - also a nonzero $\lambda_*$ is perfectly compatible with dilatation symmetry. A key argument for the existence of a fixed point $\lambda_*=0$ is stability. Indeed, a consistent fundamental theory should not have problems with stability. We recall that $\tilde U(\chi)$ should be interpreted as the effective potential with all other fields except $\chi$ and the metric integrated out, i.e. for all scalar operators taking appropriate $\chi$-dependent values. In order to ensure stability we assume that $\tilde U(\chi)$ should not diverge to minus infinity for $\chi\to\infty$. This implies the inequality
\be\label{18}
\lim_{\chi\to\infty}\lambda(\chi)=\lim_{\chi\to\infty}\frac{\tilde U(\chi)}{\chi^4}\geq 0.
\ee

It is then sufficient that $\beta_\lambda=\partial\lambda/\partial\ln\chi$ is negative for small enough positive $\lambda$. The flow for increasing $\chi$ drives $\lambda$ towards smaller values. Since $\lambda$ remains positive according to eq. \eqref{18}, the only possibility is a fixed point at $\lambda_*=0$,
\be\label{19}
\beta_\lambda(\lambda=0)=0.
\ee
Eq. \eqref{19} should hold for arbitrary values of the other couplings like $h$. If $\beta_\lambda$ is analytic in $\lambda$ we can make a Taylor expansion around $\lambda=0$ and the first term yields eq. \eqref{11}, with $A$ depending on the value of $h$ and other couplings. This form of the $\beta$-function implies for the stability matrix $S$ that the eigenvalue $A$ is typically different from the other eigenvalues. In the vicinity of the fixed point the flow of $\lambda$ decouples from the other couplings, since $\partial\beta_\lambda/\partial h^2(\lambda=0)=0$.

Of course, positive $\beta_\lambda$ for small $\lambda$ is also a logical possibility and this would imply the approach to a nonzero $\lambda_*$ for $\chi\to\infty$ or diverging $\lambda$. We should stress, however, that in presence of the condition \eqref{18} the existence of a fixed point $\lambda_*=0$ depends only on the {\em sign} of $\beta_\lambda$ for small $\lambda$. No fine tuning of parameters is involved. We also emphasize that the stability condition $V(\chi\to\infty)\geq 0$ implies in our dilatation invariant setting a positive or zero cosmological constant after Weyl scaling. In contrast, in a model without dilatation symmetry, a negative cosmological constant would not contradict the stability condition. 

\bigskip\noindent
{\em (10)\quad Which fluctuations dominate the computation of $\beta_\lambda$?}

One would like to compute the form of $\beta_\lambda$, its fixed points and the stability matrix. For this endeavor it is useful to understand first what is the dominant momentum range of the fluctuations responsible for $\beta_\lambda$. Let us compare the contribution of quantum fluctuations to the effective potential $\tilde U$ for two neighboring values $\chi_1$ and $\chi_2,~\chi_1<\chi_2$. Consider fermions which  have mass $h\chi_2$ or $h\chi_1$ for the respective values of $\chi$. The fermion mass acts as an effective infrared cutoff for the momenta of the fermion fluctuations. Changing this cutoff from $\chi_2$ to $\chi_1$ includes additional momenta in the range $(h\chi_1)^2\lesssim q^2\lesssim(h\chi_2)^2$. In consequence, the effective potential receives a contribution
\be\label{20}
\Delta\tilde U=\tilde U(\chi_2)-\tilde U(\chi_1)\approx (h\chi_2)^4-(h\chi_1)^4,
\ee
corresponding to $\Delta\lambda\sim h^4\Delta\chi/\chi$, with $\Delta\chi=\chi_2-\chi_1$. We conclude that the fermions contribute to $\beta_\lambda$ a term of the order $h^4$. For bosons the situation is similar except for the opposite sign.

In presence of many different particles the dominant contribution to $\beta_\lambda$ comes from the particles with the largest mass, i.e. the largest $h$. Typically, a unified theory will contain particles with mass of the same order as the Planck mass $\chi$. For these particles the dimensionless coupling $h$ is of the order of one and we assume that their fluctuations dominate the running of $\lambda$. As compared to them the contribution of the electron or light quark fluctuations is suppressed by more than 80 orders of magnitude and therefore completely negligible. For the fermions of the standard model the effective coupling $h$ is tiny since it is proportional to ratio between the Fermi scale and the Planck scale. 

We conclude that the properties of $\beta_\lambda$, like the location of possible fixed points and the value of the anomalous dimension $A$, involve dominantly the physics in a momentum range of the order of the Planck mass. From the point of view of the effective low energy theory described by the standard model, this momentum range concerns the ultraviolet physics. An analogue can be found in a naive calculation of the cosmological constant in the Einstein frame, where again the fluctuations  near the ultraviolet cutoff dominate. Also in the Einstein frame the variation of the cosmological constant as a function of a common scale for all particle masses would be dominated by the heaviest particles. 
As stressed before, however, the proper frame for the quantum fluctuations is the dilatation invariant Jordan frame. Only in this frame the special properties of fixed points become visible. We conclude that the issue of the role of quantum fluctuations for the cosmological constant should focus on Planck scale physics. Thus our approach differs strongly from attempts to understand the cosmological constant problem in terms of the long wavelength fluctuations, for example involving the momenta smaller than the QCD-scale $\Lambda_{QCD}$ or even the sub-eV range.

At this stage we may note an interesting property for cosmologies with approximate dilatation symmetry. For $\chi$ or $\varphi$ increasing in the course of their cosmological evolution the momentum range $q^2\approx \chi^2$ relevant for the form of the effective potential is shifted to higher values. The late universe explores the short distance physics!

\bigskip\noindent
{\em (11)\quad Does the electroweak phase transition induce a jump in the cosmological constant?}

Cosmological transitions like the electroweak transition lead to a sudden increase of the value of the Higgs field $\varphi_H$ for temperatures below some critical temperature. A similar event happens for QCD with the onset of the chiral condensate as an order parameter. A change of $\varphi_H$ is expected to contribute to the effective potential of the order $\lambda_H\varphi^4_H$, with $\lambda_H$ the quartic coupling of the Higgs scalar. This contribution alone is of the order of $\Delta U\sim (100GeV)^4$ and one may wonder if a fine tuning of the cosmological constant is needed in order to cancel it. Although both the electroweak transition and the chiral transition in QCD may be described by an analytical crossover rather than a true phase transition, the issue of a possible ``jump'' in the cosmological constant at ``late time'' (as compared to the Planck time) merits attention. How does early cosmology before these transitions ``know'' that it should prepare the value of the cosmological constant, such that a jump of its value at the transition time can be absorbed?

So far, we have only considered the effective potential $U(\varphi)$ at zero temperature. However, the early universe was in thermal equilibrium and the relevant quantity is the temperature dependent effective potential $U(\varphi,T)$ which corresponds to the free energy. One may guess that the temperature effects break the dilatation symmetry explicitely, since a particular momentum range $q^2\approx T^2$ is singled out. We will see that this question depends on the way how temperature scales under dilatation transformations. Let us demonstrate the issue in the limit where the running couplings are already close to the fixed point such that the running of dimensionless couplings can be neglected in a first approximation. 

We consider the free energy in dependence on the value of the cosmon field $\chi$ and a Higgs doublet $\chi_H$
\ba\label{21}
\tilde U(\chi,\chi_H,T)&=&\lambda_*\chi^4+
\frac{\lambda_H}{2}
(\chi^*_H\chi_H-h^2_H\chi^2)^2\nn\\
&&+\gamma\tilde T^2\chi^*_H\chi_H-\delta\tilde  T^4.
\ea
Here $\tilde T$ is the temperature in the Jordan frame, related to the temperature in the Einstein frame by $\tilde T= (\chi/M)T$ \cite{CWQ}. In a realistic universe with temperature around the Fermi scale, only the doublet is in thermal equilibrium. We have therefore not added temperature dependent terms for $\chi$, but this would not change our conclusions. We also use here an oversimplified temperature dependence corresponding to a second order phase transition - again this does not affect the outcome. For 
$\tilde T^2>\tilde T_c=\lambda_Hh^2_H\chi^2/\gamma$ the electroweak symmetry is restored, $(\langle \chi_H\rangle=0)$ whereas for $\tilde T<\tilde T_c$ spontaneous electroweak symmetry breaking occurs due to $\langle\chi_H\rangle\neq 0$. 

For a fixed $\tilde T\neq 0$ the symmetry of multiplicative rescalings of $\chi$ and $\chi_H$ is violated and therefore dilatation symmetry is explicitely broken. As a consequence, for fixed $\tilde T\neq 0$ the effective quartic coupling is no longer governed by a fixed point and obeys for $\langle\chi_h\rangle=0$
\ba\label{21A}
\lambda(T)&=&\chi^{-4}\tilde U(\chi,\langle\chi_h\rangle(\chi,\tilde T),\tilde T)\nn\\
&=&\lambda_*+\lambda_Hh^4_H/2-\delta\tilde T^4/\chi^4.
\ea
From this point of view the running of $\lambda(\chi)$ towards the fixed point $\lambda_*$ happens only for $\tilde T=0$, since only for this case the combination of dilatation symmetry and a runaway solution should end in a fixed point. This makes a vanishing temperature particular due to the larger symmetry as compared to $\tilde T\neq 0$. One would be tempted to argue that our mechanism leading to $\lambda_*=0$ only applies for the ``vacuum'' $(\tilde T=0)$, whereas the temperature fluctuations should be considered as an additional ingredient such that an apparent jump $U_{min}(\tilde T)$ between high and low $\tilde T$ needs no tuning. 

The situation is more complex, however, since we can also decide to scale the temperature $\tilde T$ by keeping the temperature in the Einstein frame $T$ fixed. With all particle masses scaling $\sim \chi$ it is a natural choice that also the temperature units scale with $\chi$. A given thermodynamic equilibrium situation corresponds then to fixed $T$, and therefore $\tilde T\sim\chi$. With this scaling the fundamental dilatation symmetry is no longer broken by temperature effects. (This contrasts to the breaking of the effective low energy dilatation symmetry by temperature effects.) In consequence, the flat valley associated to the spontaneous breaking of the fundamental dilatation symmetry persists in the presence of temperature effects. What changes, though, is the location of the flat valley. Instead of being defined by $|\chi_H|=h_H\chi$ for $T=0$, it changes to $\chi_H=0$ for $T>T_c$. As a consequence, the location of the fixed point for $\lambda$ changes to 
\be\label{22}
\lambda(T)=\lambda_*+\lambda_Hh^4_H/2-\delta T^4/M^4.
\ee
For $\lambda_*=0$ one finds a nonzero $\lambda(T)$, dominated for large $T$ by the term $\sim\delta T^4$. This reflects simply the free energy for the relativistic Higgs scalars and other particles of the standard model. Also the term $\sim \lambda_H$ should be considered as part of the free energy. In a more realistic model for the electroweak phase transition this term would have a more complicated temperature dependence - it is part of the pressure $p(T)=-\lambda_Hh^4_HM^4/2+\delta T^4$ of the standard model particles.

The stability argument in favor of a fixed point at $\lambda_*=0$ was based on the sign of the $\beta$-function for small $\lambda$ at zero temperature. Temperature effects induce a modification of the effective $\beta$-function, $\partial\ln\lambda(T)/\partial\ln\chi\neq\partial\ln\lambda/\partial\ln\chi$.
For $T\neq 0$ stability no longer requires $\lambda(T)\geq 0$. Instead, the pressure for the standard model particles should be positive. 
For large $T$ the pressure may dominate such that $\lambda(T)\approx-p(T)/M^4$ becomes negative. The stability argument for vanishing $\lambda_*$ singles out a particular temperature, namely $T=0$. This is the reason why $\lambda(T)$ vanishes precisely for $T=0$, but not for $T\neq 0$. 

We can easily translate these findings to the Einstein frame. With $\varphi_H=(M/\chi)\chi_H~,~U=(m/\chi)^4\tilde U$ we find
\be\label{23}
U=\lambda_*M^4+\frac{\lambda_H}{2}
(\varphi^*_H\varphi_H-h^2_HM^2)^2+\gamma T^2
\varphi^*_H\varphi_H-\delta T^4.
\ee
Indeed, for fixed $T$ the effective potential does not depend on $\varphi$, corresponding to the flat valley in the Goldstone direction. On the other hand, for fixed $\tilde T$ the flat valley disappears, since the replacement $T=(M/\chi)\tilde T$ induces an explicit $\chi$- or $\varphi$-dependence for the temperature effects. In principle, the choice of $\tilde T$ or $T$ is a matter of convenience since it corresponds to a choice of different variables for the free energy. Using $T$ seems, however, by far the most simple choice for most situations.

The effective potential \eqref{23} is a standard description of the temperature dependent Higgs mechanism. If we interprete the contribution $\lambda_Hh^4_HM^4/2$ as part of the cosmological constant there is indeed a ``jump'' when $|\varphi_H|$ is turned on for $T<T_c$. A better description realizes that this is part of the pressure (together with the term $\sim\delta T^4$), which jumps from a positive value to almost zero at the time of the phase transition.

\section{Dilatation symmetry in higher dimensions}
\label{dilatationsymmetryhigher}

\bigskip\noindent
{\em (12) How do higher dimensions affect the issue of dilatation symmetry?}

We have argued that the computation of the dilatation anomaly concerns dominantly the fluctuations with momenta of the order of the Planck scale or, more generally, the highest scale for masses of particles. Within higher dimensional theories, this is typically the scale below which the world looks effectively four dimensional. In terms of a characteristic compactification scale $M_c=\epsilon_c\chi=L^{-1}_c$ the higher dimensional world can be resolved for momenta larger than $M_c$, whereas observations with a long wavelength larger than $L_c$ have insufficient resolution of the additional internal dimensions, leading to an effective four dimensional description. Typically, $L_c$ is a characteristic length scale for the  geometry of internal space, including warping \cite{WH}. In case of brane worlds it may also represent the thickness of the brane, cf. sect. \ref{geometricalrunaway}. For $\epsilon_c$ independent of $\chi$ the compactification scale $M_c$ is proportional to the (four dimensional) Planck mass. 

While $M_c$ acts as an effective {\em ultraviolet} cutoff for the validity of the four dimensional description, it is an effective {\em infrared} cutoff for the validity of the higher dimensional description. From the higher dimensional point of view the issue of the dilatation anomaly concerns the infrared physics. Cosmologies with a time evolution of $M_c$ correspond to a change in the effective infrared scale of the higher dimensional theory. This change of perspective may influence our view of the problem. Within higher dimensional cosmology, the scale $M_c$ is a dynamical scale associated to the properties of cosmological solutions. Therefore, an evolution in internal geometry affects the issue of the four dimensional cosmological constant.

It is not difficult to write down a dilatation symmetric model in $d=4+D$ dimensions. We consider here gravity and a scalar field $\xi$. For a standard kinetic term, the field $\xi$ has dimension mass $^{(d-2)/2}$ such that a dilatation symmetric coupling to the higher dimensional curvature scalar $R$ is always allowed
\ba\label{24}
-{\cal L}=\sqrt{g}\left\{\frac\zeta2\partial^\mu\xi\partial_\mu\xi-\frac12\xi^2 R+F(R_{\mu\nu\sigma\lambda})\right\}.
\ea
Here $F$ contains suitable contractions of $d/2$ powers of the curvature tensor $R_{\mu\nu\sigma\lambda}$, as well as terms where one or several powers of $R_{\mu\nu\sigma\lambda}$ are replaced by pairs of covariant derivatives. This term is allowed only in even dimensions. An example is $F=\tau R^{d/2}$. For $d\neq 6$ no polynomial potential $U_d(\xi)$ is consistent with dilatation symmetry. This is a noteable difference from $d=4$ where $\xi^4$ is dilatation invariant, or $d=6$ where $\xi^3$ is allowed. (A non-canonical kinetic term, say $\xi^{d-4}\partial\xi\partial\xi$, would change the dimension of $\xi$. Now $\xi^d$ would be invariant. However, after rescaling to a standard kinetic term this would appear as a potential involving fractional powers of $\xi,U_d\sim\xi^{2d/(d-2)}$, a case that we will not consider here.) For simplicity, we restrict our discussion to $d\neq 2~ mod~4$ - otherwise additional invariants of the type $\xi R^{(d+2)/4}$ would be possible. (For $d=2~mod~4$ such terms may be forbidden by a discrete symmetry $\xi\to -\xi$.) In general, $F$ will involve dimensionless couplings like $\tau$. 

Let us first discuss
\be\label{24A}
F=\tau R^{d/2}
\ee
and address later the issue of a more general form of $F$. If our picture of a runaway solution is valid and the dimensionless couplings $\zeta$ and $\tau$ run towards fixed points, there should be asymptotic solutions with effectively constant $\zeta$ and $\tau$. We study this case first and consider the field equations for $\xi$ and $g_{\mu\nu}$
\be\label{25}
\zeta D^2\xi=-R\xi,
\ee
and
\be\label{26}
\xi^2(R_{\mu\nu}-\frac12 Rg_{\mu\nu})
=T^{(\xi)}_{\mu\nu}+T^{(\tau)}\mn.
\ee
The energy momentum tensor for $\xi$ gets modified due to the $\xi^2R$ coupling,
\be\label{27}
T^{(\xi)}\mn =\zeta\partial_\mu\xi\partial_\nu\xi-\frac{\zeta}{2}
\partial^\rho\xi\partial_\rho\xi g\mn+
D_\nu D_\mu\xi^2 -D^2\xi^2 g\mn,
\ee
where $D_\mu$ denotes a covariant derivative, wit $D^2=D^\rho D_\rho$. We have also written the contribution of the higher curvature invariant $\sim \tau$ in the form of an energy momentum tensor
\ba\label{27a}
T^{(\tau)}\mn &=&-\tau R^{\frac d2}g\mn+\tau d\left[R^{\frac d2 -1}R\mn \right.\nn\\
&&-\left(\frac d2 -1\right)R^{\frac d2 -2}(D_\nu D_\mu R-D^2 Rg\mn)\nn\\
&&-\left(\frac d2 -1\right)\left(\frac d2 -2\right)R^{\frac d2-3}\nn\\
&&\quad (D_\mu RD_\nu R-D^\rho R D_\rho R g\mn)\Big].
\ea
This is for pure convenience - we may put this piece on the l.h.s. of eq. \eqref{26} and interprete it as a modification of the gravitational part of this equation. 

Possible solutions with constant $\xi~(\partial_\mu\xi=0)$ must have $R=0$ according to eq. \eqref{25}. Then eq. \eqref{26} implies that the higher dimensional space must be an Einstein space, $R\mn-\frac 12 Rg\mn=0$. We discuss higher dimensional Einstein spaces that lead to a vanishing effective cosmological constant in four dimensions in the appendix. It is encouraging that such solutions exist - they are candidates for the asymptotic state of the universe for $t\to \infty$. However, it is not clear at this point if the universe evolves towards such a state, or towards possible other solutions with a nonvanishing four dimensional cosmological constant. We therefore keep our discussion general and concentrate on qualitative aspects. 

Even for a universe that is homogeneous in the ``normal'' three space coordinates and static, the scalar $\xi$ may depend on the internal coordinates, leading to $\partial_\alpha\xi\neq 0$, with $\alpha=1 \dots D$ denoting the internal coordinates $y^\alpha$. In this case the curvature scalar $R$ does not automatically vanish. We may contract eq. \eqref{26} with $g^{\mu\nu}$
\ba\label{28}
&&\xi^2\left(\frac d2 -1\right)R+\tau d(d-1)\left(\frac d2 -1\right)\left(R^{\frac d2-2}D^2R\right.\nn\\
&&\quad \left.+\left(\frac d2-2\right)
R^{\frac d2-3}\partial^\rho R\partial_\rho R\right)=-(T^{(\xi)})^\mu_\mu,
\ea
where
\ba\label{28A}
&&(T^{(\xi)})^\mu_\mu=-\left(\frac d2-1\right)\zeta\partial^\rho\xi\partial_\rho\xi-(d-1)D^2\xi^2\nn\\
&&=-\left[\left(\frac d2-1\right)\zeta+2(d-1)\right]
\partial^\rho\xi\partial_\rho\xi+
\frac{2(d-1)}{\zeta}\xi^2R.
\ea
In the last identity we have used the field equation \eqref{25}. A solution with $R=0$ exists only for $\partial_\rho\xi=0$. 

\bigskip\noindent
{\em (13) What determines the asymptotic value of the effective four dimensional cosmological constant?}

We are interested in a class of possible solutions, for which asymptotically all effective four dimensional scalar fields become static and homogeneous in the ordinary three space dimensions. This will include the asymptotic state of the runaway solutions discussed in sect. \ref{dilatationsymmetry}, in the sense that the kinetic energy of the scalar field vanishes for $t\to\infty$. Such solutions do not necessarily lead to a flat geometry in three space dimensions, however. For example, the asymptotic solution may be characterized by a positive effective four dimensional cosmological constant (De Sitter space) or a negative one (anti De Sitter space). We therefore look first for solutions where the four dimensional metric can differ from flat space, while the field $\xi$ depends only on the internal coordinates, but not on time and the ordinary three space coordinates. We will generalize this discussion later by allowing $\xi$ or the typical length scale of internal space $L$ to depend on time. 

Let us consider the following general ansatz for the higher dimensional metric
\be\label{29}
\hat g_{\hat \mu\hat\nu}=
\left(\begin{array}{ccc}
\sigma(y)g^{(4)}\mn(x)&,&0\\
0&,&g^{(D)}_{\alpha\beta}(y)
\end{array}\right).
\ee
From now on we denote the $d$-dimensional objects with a hat, in order to distinguish them from four dimensional objects $(\mu,\nu=0\dots 3,\hat\mu,\hat\nu =0\dots d-1,\alpha,\beta=1\dots D)$. The $d$-dimensional curvature scalar $\hat R$ involves the ``warp factor'' $\sigma(y)$ and we decompose \cite{RDW}
\ba\label{30}
\hat R&=&R^{(4)}(x)\sigma^{-1}(y)+\tilde R_D(y),\nn\\
\tilde R_D&=&R^{(D)}-\sigma^{-2}\partial^\alpha\sigma\partial_\alpha\sigma-4\sigma^{-1}D^2\sigma,
\ea
with $R^{(4)}$ and $R^{(D)}$ the curvature scalars built with the four dimensional metric $g^{(4)}\mn(x)$ and the internal metric $g^{(D)}_{\alpha\beta}(y)$, respectively. For a maximally symmetric four dimensional space the curvature scalar $R^{(4)}$ becomes independent of $x$. For a cosmological solution, it is proportional to the effective four dimensional cosmological constant.

We want to determine $R^{(4)}$ as an integral over internal space, where the integrand involves suitable combinations of $g^{(D)}_{\alpha\beta}(y),\sigma(y)$ and $\xi(y)$. For this purpose, we will follow two approaches. The first will use the field equation \eqref{25} for $\xi$. The second will employ an effective four dimensional theory, which is valid if $R^{(4)}$ is sufficiently small as compared to $M^2_c$. The combination of the two determinations will yield $R^{(4)}/M^2_c\sim \tau$. In particular, for a small enough dimensionless parameter $\tau$ the four dimensional curvature scalar turns out much smaller than $M^2_c$, such that the dimensional reduction becomes self-consistent. The validity of dimensional reduction, $R^{(4)}/M^2_c\ll 1$, extends, however, to a wide class of solutions and covers all realistic cosmologies. 

Let us first employ the field equation \eqref{25} and integrate over the higher dimensional space
\be\label{31}
\int \hat g^{1/2}\{\zeta\xi\hat D^2\xi+\xi^2\hat R\}=
\int \hat g^{1/2}\{\xi^2\hat R-\zeta\partial^{\hat \mu}\xi\partial_{\hat \mu}\xi\}=0.
\ee
This important relation can be derived independently by integration over the trace of the gravitational field equation \eqref{28}. We insert the decomposition \eqref{30} and treat $R^{(4)}$ as a constant. For $\xi$ depending only on $y^\alpha$, this implies 
\be\label{32}
R^{(4)}\int_yg^{1/2}_D\sigma\xi^2=
\int_y g^{1/2}_D\sigma^2\{\zeta\partial^\alpha\xi\partial_\alpha\xi-\xi^2\tilde R_D\},
\ee
where $g_D=\det (g^{(D)}_{\alpha\beta})$. Indeed, this determines $R^{(4)}$ in terms of suitable $y$-integrals, independently of the details of the full solution for the field equations. In particular, the identity \eqref{32} does not involve $\tau$. It will remain valid for a more general function $F$ in eq. \eqref{24}, provided that this function does not depend on $\xi$.

Next we proceed to dimensional reduction. This is achieved by integrating out the internal dimensions. In principle, the reduced four dimensional action contains infinitely many fields, but we are interested here only in the four-dimensional metric $g^{(4)}\mn(x)$. For small enough $R^{(4)}$ we expand up to terms linear in $R^{(4)}$ 
\ba\label{33}
-S&=&\int_xg^{1/2}_4(x)\int_y g^{1/2}_D(y)\sigma^2(y)\big[A_1(y)\nn\\
&&-\frac12 A_2(y)R^{(4)}(x)+\dots \big],
\ea
with
\ba\label{34}
A_1&=&\frac{\zeta}{2}\partial^\alpha\xi\partial_\alpha\xi-\frac12\xi^2\tilde R_D+\tau(\tilde R_D)^{d/2},\nn\\
A_2&=&\sigma^{-1}\left[\xi^2-\tau d(\tilde R_D)^{\frac d2-1}\right].
\ea
The coefficient of the term linear in $R^{(4)}$ can be identified with the squared effective four dimensional Planck mass $\chi$,
\be\label{35}
\chi^2=\int_yg^{1/2}_D\sigma^2A_2.
\ee
Similarly, we infer the effective cosmological constant or, more generally, the effective potential $\tilde U(\chi)$ as
\be\label{36}
\tilde U(\chi)=\int_y g^{1/2}_D\sigma^2 A_1.
\ee
We note that the first two pieces in $A_1$ contribute to the integral a structure exactly equal to the r.h.s. of eq. \eqref{32}.

In the limit where $\partial_\mu\chi$ can be neglected, the effective four dimensional field equation reads
\be\label{37}
\chi^2(R^{(4)}\mn-\frac12 R^{(4)}g^{(4)}\mn)=-\tilde U(\chi)g\4\mn.
\ee
This is indeed equivalent to the higher dimensional field equations, provided that all  fields except the scalars and $g\4\mn$ have vanishing values (e.g. spin-one fields and massive ``Kaluza-Klein-gravitons''), and the scalar fields are (approximately) static and homogeneous. One infers from eq. \eqref{37} a second relation for $R\4$, namely
\be\label{38}
R\4=\frac{4}{\chi^2}\tilde U(\chi).
\ee

We can now combine eq. \eqref{38} with the field eq. for $\xi$, i.e. eq. \eqref{32}. The latter can be written in the form 
\be\label{39}
R\4=\frac{2}{\mu_\xi}\tilde U_\xi~,~\mu_\xi=\int_y g^{1/2}_D\sigma\xi^2,
\ee
where we define
\ba\label{40}
\tilde U_\xi&=&\int_y g^{1/2}_D\sigma ^2\left\{\frac{\zeta}{2}\partial^\alpha\xi\partial_\alpha\xi-\frac12\xi^2\tilde R_D,\right\},\nn\\
\tilde U_\tau&=&\tau\int_y g_D^{1/2}\sigma^2(\tilde R_D)^{\frac d2},
\ea
such that $\tilde U=\tilde U_\xi+\tilde U_\tau$. We also use
\be\label{42}
\chi^2=\mu_\xi+\mu_\tau~,~\mu_\tau=-\tau d\int_y g^{1/2}_D\sigma(\tilde R_D)^{\frac{d}{2}-1}.
\ee
Writing eq. \eqref{38} as 
\be\label{40A}
R\4=\frac{4}{\mu_\tau+\mu_\xi}
(\tilde U_\xi+\tilde U_\tau),
\ee
we conclude that the solution $\xi(y)$ always adjusts itself such that 
\be\label{41}
\tilde U_\xi=2
\left(\frac{\mu_\tau}{\mu_\xi}-1\right)^{-1}\tilde U_\tau.
\ee
In consequence, the effective potential is proportional to $\tau$, 
\be\label{43}
\tilde U=\tilde U_\xi+\tilde U_\tau=
\frac{\mu_\tau+\mu_\xi}{\mu_\tau-\mu_\xi}~\tilde U_\tau
\ee
This answers our question: the asymptotic value of the effective four dimensional constant $\tilde U$ is proportional to $\tau$ and to an integral over $(\tilde R_D)^{d/2}$. It vanishes for $\tau =0$ or $\tilde R_D=0$. 

We note $\mu_\xi>0$ and we will choose $\tau>0$ such that for $d=4 ~mod ~4$ one infers $\tilde U_\tau\geq 0$. The sign of $\tilde R_D$ is a priori not fixed. (Our conventions imply a positive $R^{(D)}$ for a sphere.) Any realistic cosmology requires a positive gravitational ``constant'', $\chi^2>0$ or $\mu_\tau>-\mu_\xi$. A positive effective cosmological constant, $\tilde U>0~,~R\4>0$ requires by eq. \eqref{39} $\tilde U_\xi>0$. Such solutions can only exist for $\mu_\tau>\mu_\xi$ (cf. eq. \eqref{41}. In turn, a negative cosmological constant, $\tilde U<0~,~R\4<0$ can only be realized for $\mu_\tau<\mu_\xi$. Solutions with an asymptotically vanishing cosmological constant, $\tilde U=0$, require $\tilde U_\tau=0$, and therefore $\tilde U_\xi=0$. No restriction on $\tilde \mu_\tau$ (except $\tilde \mu_\tau>-\tilde \mu_\xi$) arises in this case. 

Special solutions are those with $\xi=const.$. These are the same solutions as for a model without a dilaton field, where $\xi$ can be regarded as a coupling constant that violates dilatation symmetry explicitely. As discussed above, the solutions for $\xi\neq 0$ are $d$-dimensional Einstein spaces. For this case a class of warping solutions of the type \eqref{29} has been discussed in \cite{RDW}. One finds solutions with arbitrary $\tilde U$, including $\tilde U=0$. The static solutions with a vanishing four dimensional cosmological constant, $\lambda M^4=\tilde UM^4/\chi^4=0$, are discussed in the appendix. (For $\xi=0$ additional solutions exist since $\hat R=0$ is sufficient.) We will remain here more general and include the possibility that $\partial_\alpha\xi\neq 0$. 

We will discuss later solutions where both $\tilde U_\tau$ and $\mu_\tau/\mu_\xi$ vanish asymptotically. Eqs. \eqref{38} and \eqref{40A} can be obeyed simultaneously only for $\tilde U_\xi=0$, and therefore a vanishing cosmological constant $\tilde U=0$. We will see that this is indeed the case for these solutions. For the approach to this asymptotic solution, eq. \eqref{43} may suggest a negative $\tilde U$. We point out, however, that our discussion only holds for the asymptotic state, where terms $\sim \partial_\mu\chi$ and possible contributions to the field equations from incoherent matter and radiation can be neglected. As a consequence, eq. \eqref{41} cannot be used for the approach to the asymptotic state and should be replaced by a more general relation 
\be\label{48A}
\tilde U_\xi=Q\tilde U_\tau~,~\tilde U=(1+Q)\tilde U_\tau,
\ee
with $Q$ a constant depending on the particular approach to the asymptotic state. This is justified whenever the contributions neglected for our discussion of the asymptotic state are of the same order as $\tilde U$. If they a smaller, $Q$ is given by eq. \eqref{41}. If they are much larger, our discussion of the asymptotic solution may need to be modified. The property \eqref{48A} will be important for our later discussion, since it implies that the $\xi$-dependent contribution $\tilde U_\xi$ does not induce a characteristic scale that is very different from the pure geometrical contribution $\tilde U_\tau$. The scale for $\tilde U$ is only set by $\tilde R_D$. 

\bigskip\noindent
{\em (14) How does the compactification scale influence the asymptotic cosmological constant?}

The geometry of internal space typically involves a characteristic length scale $L=M^{-1}_c$. We want to understand how the effective potential $\tilde U$ depends on this scale. For this purpose we write $g^{(D)}_{\alpha\beta}=g^{D,0}_{\alpha\beta}(y)L^2$ and we consider the dependence of the four dimensional effective action on $L$, while keeping $g^{D,0}_{\alpha\beta},\sigma$ and $\xi$ fixed. If $L$ is allowed to depend on $x$ it will appear as a four-dimensional scalar field, often called the ``radion''. Similarly, we may make an ansatz $\xi=\xi(x)\xi^{(0)}(y)$ with a fixed reference function $\xi^{(0)}(y)$, which is typically proportional to a solution of the field equations. Again, $\xi(x)$ is a four-dimensional scalar field. The four dimensional effective action \eqref{33}, with eqs. \eqref{35}, \eqref{36}, allows us to interprete $\tilde U$ as the effective potential for the scalars $L$ and $\xi$, and to study the dependence of the effective Planck mass $\chi$ on these scalars. A computation of the kinetic terms of the scalars requires to introduce $L(x)$ explicitely in the ansatz \eqref{29} and to take terms $\sim \partial_\mu \xi$ into account. With $g^{1/2}_D\sim L^D~,~\tilde R_D\sim L^{-2}$ we infer 
\be\label{44}
\tilde U=c_1L^{D-2}\xi^2+\tau c_2L^{-4},
\ee
where $c_1$ and $c_2$ depend on the precise shape of internal geometry, the warping function $\sigma$ and the solutions $\xi^{0)}(y)$. 

We will distinguish between two different types of solutions, the ``scaling solutions'' where $\xi^2L^{D+2}$ is constant, and the ``geometrical runaway solutions'', where $\xi^2L^{D+2}$ diverges asymptotically. For the scaling solutions, $\xi$ scales with $M_c$ according to its canonical dimension, $\xi\sim M_c^{\frac{d-2}{2}}$. For constant $c_1,c_2 >0$, we may compute the minimum of $\tilde U(L)$ for fixed $\xi$. This ``valley'' is characterized by
\be\label{45}
\xi^2_v=G_0\tau L^{-(D+2)}~,~G_0=\frac{4}{D-2}\frac{c_2\tau}{c_1},
\ee
We have displayed the dependence on $\tau$ explicitely in order to see that a multiplicative rescaling of $\tau$ comes in pair with a similar rescaling of $\xi^2$. in fact, this holds for all solutions derived from the action \eqref{24}, \eqref{24A}, since a common rescaling of $\tau$ and $\xi^2$ only results in an overall multiplicative factor for the action and does not affect the field equations. The valley \eqref{45} is, however, not the true trajectory of the solutions. The couplings of $\xi$ and $L$ to the four dimensional curvature scalar $R^{(4)}$ modify the exact location of the valley. The latter can only be computed in the Einstein frame and will lead to a different value of $G_0$. We discuss the determination of $G_0$ in the next section, both for constant $c_i$ and for $c_i$ depending on $L$ and $\xi$.

Inserting eq. \eqref{45} one finds the effective potential along the valley
\ba\label{46}
\tilde U_v=L^{-4}(c_1G_0+c_2\tau).
\ea
For constant $c_1,c_2$ and $G_0$ the effective potential $\tilde U_v$ vanishes for $L\to \infty$. One may therefore expect that cosmological solutions where $L\to\infty$ lead to an effectively vanishing cosmologocal constant. This does not necessarily hold, however, in our dilatation invariant setting. The crucial quantity is $\lambda=\tilde U/\chi^4$, and we need the dependence of $\chi$ on $L$, 
\be\label{48}
\chi^2=c_3L^D\xi^2+\tau c_4L^{-2},~\chi^2_v=
\left(c_3G_0+c_4\tau\right)L^{-2},
\ee
where the second identity in eq. \eqref{48} holds for the scaling solution. For the solutions of the type \eqref{45} one finds that $\lambda_*$ becomes independent of $L$. For $c_1=0$ it is proportional $\tau$. We conclude that for an exact higher dimensional dilatation symmetry the valley solutions typically lead to $\lambda_*\neq 0$ and therefore to a non-vanishing four dimensional cosmological constant. We will see in sect. \ref{dilatationinhigher} how a higher dimensional dilatation anomaly may induce $\tau\to 0$ and therefore $\lambda\to 0$. 

A second class of solutions are the geometrical runaway solutions, where $\xi^2 L^{D+2}$ diverges asymptotically. A particular realization corresponds to constant $\xi$ and increasing $L\to\infty$. Then the effective four dimensional cosmological constant indeed vanishes asymptotically for $L\to\infty$
\be\label{49}
\lambda =\frac{c_1}{c^2_3}\xi^{-2}L^{-(D+2)}=\frac{c_1}{c_3}(L\chi)^{-2}.
\ee
Such solutions lead generically to an asymptotically vanishing cosmological constant.

\bigskip\noindent
{\em (15) Can geometrical runaway solutions solve the cosmological constant problem?}

From the cosmological point of view, the geometrical runaway solutions, $\xi(t)\to~const$, $L(t)\to\infty$, are perfectly valid solutions of the cosmological constant problem. Indeed, solutions of this type have been discussed in detail \cite{SchW} in the context of a six dimensional gauge theory. The asymptotic vanishing of the cosmological constant is simply due to the fact that $\chi^4$ increases with a higher power of $L$ than $\tilde U$. For constant $\xi$ this is rather generic, since $\chi^2$ increases $\sim L^D$, and $\tilde U$ increases at most $\sim L^D$. (In our dilatation symmetric setting without a higher dimensional potential $V(\xi)$, the increase of $\tilde U$ is actually only $\sim L^{D-2}$.) We note that geometrical runaway solutions can also arise in a more general context, where $\xi$ changes but decreases less fast than $L^{-\frac{D+2}{2}}$. (The scaling solutions correspond to the boundary where no geometrical runaway occurs anymore.) In our model, the term $\sim \tau R^{\frac d2}$ becomes asymptotically irrelevant for the geometrical runaway solutions, as far as the contribution to $\chi^2$ is concerned. It may influence, however, the geometry via $g^{D,0}_{\alpha\beta}$ and $\chi^{(0)}$. 

For a geometrical runaway solution, the value of $L$ at late times is large as compared to $\xi^{-\frac{2}{D+2}}$, but not yet infinite. In this sense the geometrical runaway solutions are models of large extra dimensions \cite{LED}, where a large effective Planck mass is associated to a large size of internal space. Such models with large extra dimensions have been noted already among the first realistic Kaluza-Klein cosmologies \cite{CWKK}. 

In a four dimensional description, the geometrical runaway typically leads to an effective potential that vanishes exponentially with $\varphi$, e.g. eq. \eqref{1} with $\lambda_*=0$. In the dimensionally reduced effective action the kinetic term for $L$ appears in the form
\be\label{50}
-{\cal L}_{kin}=\frac12(c_5\xi^2L^{D-2}+\tau c_6 L^{-2})
\partial_\mu L\partial^\mu L,
\ee
where the term $\sim \tau$ becomes negligible for $L\to\infty$. After Weyl scaling, this term gets multiplied by $M^2/\chi^2$, such that in the Einstein frame
\ba\label{51}
&&-{\cal L}^{(EF)}_{kin}\sim M^2L^{-2}\partial^\mu L\partial_\mu L~,~\nn\\
&&U=\frac{M^4}{\chi^4}\tilde U=\frac{c_1}{c^2_3}
\frac{M^4}{\xi^2}L^{-(D+2)}.
\ea
Therefore the field with canonical kinetic term reads, with $L_0=\xi^{-2/(D+2)}$, and for constant $c_1,c_3$, 
\be\label{52}
\varphi=\kappa M\ln(L/L_0)~,~U\sim M^4\exp
\left\{-\frac{(D+2)}{\kappa}\frac\varphi M\right\}.
\ee
We find indeed an exponential potential $U(\varphi)$, with $\lambda_*=0$. 

We can  cast the physics of the geometrical runaway solution into the form of a renormalization group equation for the $\chi$-dependence of the coupling $\lambda=\tilde U/\chi^4$, namely
\ba\label{53}
\frac{\partial\lambda}{\partial\ln\chi}&=&-4\lambda+\chi^{-4}
\left[\frac{\partial\tilde U}{\partial\ln L}\frac{\partial\ln L}{\partial\ln \chi}+
\frac{\partial\tilde U}{\partial \ln \xi}\frac{\partial\ln \xi}{\partial\ln\chi}\right]\nn\\
&=&\left[-4+(D-2)\frac{\partial\ln L}{\partial\ln \chi}+2\frac{\partial\ln\xi}{\partial\ln\chi}\right]\lambda.
\ea
Here we have neglected the contribution $\sim\tau$ in eq. \eqref{44}. We note that for a variation of $\chi$ at constant $\xi$ one has $\partial\ln\xi/\partial\ln\chi=0$ and $\partial\ln L/\partial\ln \chi=2/D$ (cf. eq. \eqref{48} with $\tau=0$). This is indeed of the form \eqref{10}, \eqref{11}
\be\label{54}
\frac{\partial\lambda}{\partial\ln\chi}=-A\lambda~,~A=2+\frac{4}{D}.
\ee
Geometrical runaway naturally realizes both a positive anomalous dimension $A>0$ and a fixed point value $\lambda_*$.

The presence of an anomalous dimension $A>0$ may appear as a surprise, since the effective action \eqref{24} is fully dilatation symmetric. For the geometrical runaway solutions the fundamental dilatation symmetry is spontaneously broken by a constant value of $\xi$. Since the effective four dimensional gravitational constant is given by the field $\chi$, we may formulate an effective four dimensional dilatation symmetry in the sector of gravity and $\chi$. It is broken by the potential $\tilde U(\chi)$, but restored in the limit $\chi\to\infty$ where $\tilde U\to 0$. We emphasize that the fundamental dilatation symmetry plays no essential role for the geometrical runaway - this occurs equally for models where $\xi$ is set as a constant in a higher dimensional action witout dilatation symmetry.

The reason, why geometrical runaway solutions with large extra dimensions have not been investigated more intensively in the past, is related to the particle physics aspects of such a solution to the cosmological constant problem. Indeed, while a fixed point $\lambda_*=0$ is realized naturally, a non-trivial fixed point for the gauge couplings or the Yukawa couplings of the standard model is less obvious. As we have discussed in sect. \ref{timevarying}, realistic particle physics and cosmology need $h_*\neq 0,\lambda_*=0$. In the simplest setting one would associate $L^{-1}$ with the unification scale $M_{GUT}$ of a grand unified model. Then the geometrical runaway would lead to a strong decrease of $M_{GUT}/M_p\sim(L\chi)^{-1}\sim L^{-(D+1)}$, much stronger than allowed by nucleosynthesis \cite{DSW}. Similar problems may be expected for a time dependence of the Fermi scale of weak interactions. However, already the first ideas about quintessence have emphasized that for any realistic cosmology the change in the ratio $M_W/M_p$ must have been small for all epochs after nucleosynthesis \cite{CWQ1}. The mass of the $W$-boson $M_W$ must scale proportional to $\chi$ for any realistic cosmology. This seems far from obvious in simple realizations of electroweak symmetry breaking in a higher dimensional context. 

Finally, if the gauge symmetries of the standard model are realized as isometries of internal space, the effective gauge coupling typically shows a strong dependence on $L$ \cite{SchW}. This also holds for simple internal geometries if the four dimensional gauge symmetries are part of the gauge symmetries in a higher dimensional model \cite{SchW}. A strong dependence of the gauge couplings on time is not compatible with observation. We conclude that the problems of geometrical runaway are not the understanding of an asymptotically vanishing cosmological constant - this is solved naturally. They arise from the difficulty of realizing non-trivial fixed points for the dimensionless couplings and mass ratios in the standard model. We will turn to a possible solution of this problem in sect. \ref{geometricalrunaway}.

A second potential problem is connected with the size of internal space at the present epoch. The present value of $\chi$ is given by the (reduced) Planck mass, $\chi(t_0)=M\approx 10^{18}$ GeV, and observation requires for the present value of the cosmological constant $\lambda(t_0)\approx 10^{-120}$. From eq. \eqref{49} one infers
\be\label{55}
L(t_0)M=\sqrt{\frac{c_1}{c_3\lambda(t_0)}}\approx\sqrt{\frac{c_1}{c_3}}10^{60}.
\ee
For $c_1$ of the same order as $c_3$ this implies that the present characteristic size of the internal dimensions is comparable to the present horizon $H(t_0)^{-1}$. This means that they are not small and gravity in the solar system would look multi-dimensional, obviously in contradiction with our everyday experience with three space dimesions. We note that the estimate \eqref{49} and therefore eq. \eqref{55} are independent of the assumption of constant $\xi$ - they hold for all geometrical runaway solutions with arbitrary $\xi(L)$.

A possible solution to this problem is $c_1=0$, or $c_1$ depending on $L$ such that the power in eq. \eqref{49} is changed to
\be\label{56}
\lambda\sim (L\chi)^{-K}.
\ee
For sufficiently large $K$ the present length scale for internal space is indeed small enough in order to escape detection with present experiments. From
\be\label{57}
L(t_0)M\approx 10^{120/K}
\ee
one finds for $K\geq 4$ an ``internal size'' $L(t_0)\leq 10^3 eV^{-1}$. For constant $\xi$ an analysis similar to eqs. \eqref{51}-\eqref{54} replaces $D+2$ by $K$. We obtain again an exponential cosmon potential, with 
\be\label{62Aa}
A=\frac{2K}{K-2}.
\ee

At this point the field equation for $\xi$ plays a crucial role in our dilatation symmetric setting. Combining the identity \eqref{31} with dimensional reduction, we have obtained the general relation $\tilde U_\xi=Q\tilde U_\tau$, with (approximately) constant $Q$. This yields
\be\label{62A}
c_1=\frac{\tau}{\xi^2 L^{D+2}}Qc_2~,~\tilde U=\tau c_2(1+Q)L^{-4}.
\ee
The self-adjustment of $\xi^{(0)}(y)$ according to eq. \eqref{31} indeed implies a vanishing of $c_1$ for $L\to\infty$. For constant $\xi,\tau,c_2,Q$ we infer $K=2D+4$. For $D=6(12)$ this implies a small internal space at present
\be\label{62B}
L_0^{-1}\approx 10^{-2}(10^8)GeV,
\ee
such that only the problem with time varying couplings remains for the geometrical runaway solutions. The associated scale of the effective higher dimensional Planck mass is
\be\label{65A}
\xi^{\frac{2}{D+2}}=10^{-\frac{60D}{(D+2)^2}}M\approx 10^{12}(10^{15})GeV.
\ee

\section{Simple Model with vanishing cosmological constant}
\label{simplemmodel}
Let us consider the model with $\tau=0$ or $F=0$ in eq. \eqref{24}. We will see below that this situation is approached dynamically for several of our scenarios. The field equations simplify considerably
\ba\label{S1}
\hat D^2\xi&=&-\frac 1\zeta\hat R\xi,\nn\\
\hat R\hmn -\frac12\hat R\hat g\hmn&=&\frac{1}{\xi^2}\left(\zeta\partial_{\hat\mu}\xi\partial_{\hat\nu}\xi-
\frac{\zeta}{2}\partial^{\hat\rho}\xi\partial_{\hat\rho}\xi\hat g\hmn\right.\nn\\
&&\left.+D_{\hat\nu}D_{\hat\mu}\xi^2-\hat D^2\xi^2 \hat g\hmn\right).
\ea
Taking the trace of the second equation
\be\label{S2}
\frac{\hat R}{\zeta}=\frac{1}{\xi^2}\left\{\partial^{\hat\rho}\xi\partial_{\hat\rho}\xi+
\frac{2(d-1)}{(d-2)\zeta}\hat D^2\xi^2\right\},
\ee
and inserting into the field equation for $\xi$, yields
\ba\label{S3}
\xi\hat D^2\xi+\partial^{\hat\rho}\xi\partial_{\hat\rho}\xi&=&0~,~\hat D^2\xi^2=0,\nn\\
\partial_{\hat\rho}\{\hat g^{1/2}\xi\partial^{\hat\rho}\xi)&=&0~,~\xi^2\hat R=\zeta\partial^{\hat\rho}\xi\partial_{\hat\rho}\xi.
\ea
We conclude that for any solution of the higher dimensional field equations, the Lagrangian and the action vanish. Using the decomposition \eqref{29},\eqref{30} one finds
\be\label{S4}
\sigma^{-1}(R^{(4)}-\zeta\partial^\mu\ln\xi\partial_\mu\ln\xi)=-
\tilde R_D+\zeta\partial^\alpha\ln\xi\partial_\alpha\ln\xi.
\ee

We may perform dimensional reduction and find for small enough $R^{(4)},\partial^\mu\xi\partial_\mu\xi$ and $\partial_\mu L\partial^\mu L$ the effective four dimensional action $S=\int_x(g^{(4)})^{1/2}{\cal L}$ 
\ba\label{S5}
-{\cal L}&=&c_1\xi^2L^{D-2}+\frac12 c_3L^D
\{\partial^\mu\xi\partial_\mu\xi-\xi^2 R^{(4)}\}\nn\\
&&+\frac12 c_5\xi^2L^{D-2}\partial^\mu L\partial_\mu L.
\ea
Here we have made again the ansatz $\xi(\hat x)=\xi^{(0)}(y)\xi(x)$ and we generalize eq. \eqref{29} by $\hat g_{\alpha\beta}=L^2(x)g^{(0)}_{\alpha\beta}(y)$. We want to evaluate
\be\label{S6}
c_1=\frac12\int_y\sigma^2(g^{(0)}_D)^{1/2}(\zeta\partial^\alpha\xi^{(0)}\partial_\alpha\xi^{(0)}-
\xi^{(0)2}\tilde R^{(0)}_D),
\ee
where $\tilde R^{(0)}_D$ is computed with $g^{(0)}_{\alpha\beta}$, and in eq. \eqref{S6} this metric is also used to raise and lower internal indices.

We have to specify the criteria for the selection of the functions $\sigma(y),g^{(0)}_{\alpha\beta}(y)$ and $\xi^{(0)}(y)$, for which $c_1,c_3,c_5$ are computed. The dimensionally reduced action \eqref{S5} may be regarded as an expansion of the higher dimensional action in the small quantities $R^{(4)},\partial^\mu\xi\partial_\mu\xi$ and $\partial^\mu L\partial_\mu L$, where the integration over the internal coordinates $y$ is performed. We therefore expand around some metric of the type \eqref{29}, where $g^{(4)}\mn=\eta\mn$, and $L(x)$ takes a constant value. Also $\xi(y,x)=\xi^{(0)}(y)\xi$ is taken independent of $x$ with arbitrary constant $\xi$. Our criterion for the selection of $\sigma,g^{(0)}_{\alpha\beta}$ and $\xi^{(0}$ is now that these functions should correspond to a solution of the higher dimensional field equations with $R^{(4)}=0,\xi(x)=\xi,L(x)=L$. Such solutions indeed exist - for the special case $\xi^{(0)}(y)=const.$ they are discussed in the appendix.

For a computation of $\tilde R^{(0}_D$ we may use the $\alpha\beta$-component of the second field equation \eqref{S1}
\be\label{S7}
\hat R_{\alpha\beta}-\frac12\hat R\hat g_{\alpha\beta}=\frac{1}{\xi^2}
\left(\zeta\partial_\alpha\xi\partial_\beta\xi-\frac12\xi^2\hat R\hat g_{\alpha\beta}-D_\beta D_\alpha\xi^2\right).
\ee
Contracting with $\hat g^{\beta\alpha}$, we obtain
\ba\label{S8}
\hat R_{\alpha\beta}\hat g^{\beta\alpha}&=&\frac{\zeta}{\xi^2}
\partial^\alpha\xi\partial_\alpha\xi-\frac{1}{\xi^2}D^\alpha D_\alpha\xi^2\nn\\
&=&\zeta\partial^\alpha\ln\xi\partial_\alpha\ln\xi+\frac{1}{\xi^2}D^\mu D_\mu\xi^2.
\ea
Here we have also used eq. \eqref{S3}. We can next employ $\partial_\mu\xi=0$. For the solutions with $R^{(4)}=0$ one finds $\hat R_{\alpha\beta}\hat g^{\beta\alpha}=\tilde R_D$ and therefore
\be\label{S9}
\tilde R^{(0)}_D=\zeta\partial_\beta\ln\xi^{(0)}\partial_\alpha\ln\xi^{(0)}g^{(0)\alpha\beta}.
\ee
This yields $c_1=0$, such that the effective four dimensional Lagrangian \eqref{S5} contains only derivative terms. 

In other words, expanding around an internal geometry where $g^{(0)}_{\alpha\beta},\sigma$ and $\xi^{(0)}$ correspond to a higher dimensional solution with vanishing $R^{(4)}$, leads to a consistent dimensional reduction. The effective four dimensional theory indeed admits solutions where $R^{(4)}=0,~\xi(x)=const,~L=const.$, such that the effective four dimensional cosmological constant vanishes. This may seem trivial, since we have expanded around a higher dimensional solution with flat four dimensional space. A closer inspection shows, however, that the property of a consistent dimensional reduction, combined with static $\xi$ and $L$, singles out the cosmologies with a vanishing cosmological constant. In order to demonstrate this, we also expand around possible neighboring geometries that correspond to higher dimensional solutions with a nonvanishing constant $\bar R^{(4)}\neq 0$, with $\bar \xi$ and $\bar L$ independent of $x$. Dimensional reduction for such geometries leads to a nonvanishing constant $c_1$,
\be\label{74A}
c_1=\bar R^{(4)}c_3.
\ee
We will wee that dimensional reduction for such geometries is not consistent with static $\xi$ and $L$.

For this purpose we solve the field equations that follow from the variation of the four dimensional action with respect to $g^{(4)}\mn,\xi(x),L(x)$. It is convenient to perform a Weyl scaling
\be\label{S10}
g^{(4)}\mn=\frac{M^2}{L^D\xi^2c_3}g\mn.
\ee
In terms of the metric $g\mn$ in the Einstein frame, the four dimensional action becomes
\ba\label{S11}
-S&=&\int_xg^{1/2}
\left\{-\frac12 M^2R+\frac{M^2}{2}[\partial^\mu\ln\xi\partial_\mu\ln\xi\right.\nn\\
&&+\frac{c_5}{c_3}\partial^\mu\ln L\partial_\mu\ln L+6\partial^\mu\ln(\xi L^{D/2})\partial_\mu\ln(\xi L^{D/2}]\nn\\
&&\left.+\frac{c_1M^4}{c^2_3}\xi^{-2}L^{-(D+2)}\right\}.
\ea
We may introduce the ``radion''
\be\label{S12}
\varphi=\gamma_\varphi\ln(\xi L^{\frac{D+2}{2}})
\ee
and a dilaton type field
\be\label{S13}
\delta=\gamma_\delta M\ln(\xi L^\eta)
\ee
with $\eta,\gamma_\phi,\gamma_\delta$ chosen such that the kinetic terms of $\varphi$ and $\delta$ are orthogonal and normalized,
\ba\label{S14}
-S&=&\int_x g^{1/2}\left\{\frac12[\partial^\mu\varphi\partial_\mu\varphi+\partial^\mu\delta\partial_\mu\delta-M^2R]\right.\nn\\
&&\left.+\frac{c_1M^4}{c^2_3}\exp\left(-\frac{2\varphi}{\gamma_\phi M}\right)
\right\}.
\ea

For $c_1=0$ the late cosmology has attractor solutions for which $\varphi$ and $\delta$ settle to constant values. (These values depend on the initial conditions.) We end with the Einstein equations with a vanishing four dimensional cosmological constant! We conclude that the solutions with an asymptotically vanishing four dimensional curvature scalar $R=0,~R^{(4)}=0$ are compatible with dimensional reduction. 

This coincides with the simple observation that for an effective cosmological constant $\lambda_*M^4$ in the Einstein frame, the solutions of the field equation
\be\label{S14a}
R\mn-\frac12 Rg\mn=-\lambda_*M^2 g\mn
\ee
obey $R=4\lambda_*M^2$. Reinserting into the action, this would yield
\be\label{S15}
-S=\int_x g^{1/2}\left(\lambda_*M^4-\frac{M^2}{2}R\right)=-\int_xg^{1/2}\lambda^*M^4.
\ee
Only $\lambda_*=0$ is compatible with the observation that $S$ vanishes for all solutions of the higher dimensional field equations, as mentioned after eq. \eqref{S3}.

What happens with possible higher dimensional solutions where $\bar R^{(4)}\neq 0,~c_1\neq 0$? After dimensional reduction and in the Einstein frame, they lead to solutions where $\varphi$ is not static, but rather evolves asymptotically to infinity, while a static $\delta$ is allowed. In consequence, the curvature scalar $R$ in the Einstein frame and the effective cosmological constant vanish asymptotically. At any finite time, however, $R,~U(\varphi)$ and $\partial^\mu\varphi\partial_\mu\varphi$ differ from zero. The curvature tensor in the Jordan frame $R^{(4)}$ receives contributions from $R$ and from kinetic energy associated to the evolution of $\xi$ and $L$
\ba\label{81A}
R^{(4)}=\frac{c_3}{M^2}\xi^2L^D
\Big\{R&-&6\partial^\mu\ln(\xi L^{D/2})\partial_\mu\ln(\xi L^{\frac D2})\nn\\
&+&6 D^\mu D_\mu\ln(\xi L^{D/2})\Big\}.
\ea
A nonvanishing $R^{(4)}$ can be consistent with an asymptotically vanishing $R$. The precise relation of 
four dimensional solutions for $c_1\neq 0$ and $\varphi$ moving to infinity and possible higher dimensional solutions remains to be clarified. We concentrate on the solutions with $c_1=0$ for which dimensional reduction is fully consistent with static $\xi$ and $L$. 

For $c_1=0$, both $\varphi$ and $\delta$ correspond to flat directions or valleys. The flat direction for $\delta$ persists as long as dilatation symmetry is exact, while a higher dimensional dilatation anomaly will induce a potential for $\delta$. This will be discussed in the next section. On the other hand, the flat direction for the radion field is a property of the very simple action with $\tau=0$. A nonzero coupling $\tau$ will typically induce a potential for the radion. Now, an additional contribution to the potential arises from dimensional reduction of the term $\tau \hat R^{d/2}$, and also $c_1=0$ is no longer a solution. We will discuss this more general setting in the following. We will also see how a dilatation anomaly may drive $\tau$ towards zero, such that asymptotically the simple action with $\tau=0$ becomes relevant.

\section{Dilatation anomaly in higher dimensions}
\label{dilatationinhigher}

A higher dimensional dilatation anomaly can be associated to running dimensionless couplings. In our model, the parameters $\zeta$ and $\tau$ may depend on $\xi$. Let us consider a constant $\zeta$ and a running of $\tau$ according to an anomalous dimension
\be\label{58}
\frac{\partial}{\partial \ln\xi}\tau=-A_\tau \tau.
\ee
The resulting behavior
\be\label{59}
\tau=\tau_0\left(\frac \xi\mu_\tau\right)^{-A_\tau}
\ee
can lead to an asymptotic vanishing of $\tau$. This makes such models interesting candidates for an asymptotically vanishing dark energy. 

For a $\xi$-dependent coupling $\tau(\xi)$ the higher dimensional field equations \eqref{25}, \eqref{26} get modified. Nevertheless, the qualitative behavior of cosmological solutions can be understood from the properties of the four-dimensional effective potential in the Einstein frame
\be\label{60}
U=\frac{M^4}{\chi^4}\tilde U=M^4\lambda={M^4}{\tau(\xi)}
\frac{c_1L^{D+2}\xi^2/\tau(\xi)+c_2}{(c_3L^{D+2}\xi^2+c_4\tau(\xi))^2}.
\ee
We may perform a change of variables and write
\ba\label{61}
&&U(\tau,G)/M^4=\tau\frac{c_1G/\tau+c_2}{(c_3G+c_4\tau)^2}
=\tau\frac{(1+Q)c_2}{(c_3G+c_4\tau)^2},
\nn\\
&&G=L^{D+2}\xi^2.
\ea
Valley solutions are realized asymptotically if, for a fixed $\tau,~U(G)$ has a minimum at $G_0$. Then $L(\xi)=(G_0/\xi^2)^{1/(D+2)}$ defines the valley. Along the valley the effective potential
\be\label{62}
U_v(\tau)={U_0}{\tau}~,~U_0=\frac{U(\tau, G_0)}{\tau}
\ee
decreases to zero as $\tau$ goes to zero. The presence of $U_v$ indeed drives $\xi$ into the direction for which $\tau$ decreases, thus leading to asymptotically vanishing dark energy for $t\to\infty$. Furthermore, if $G_0,c_3,c_4$ depend only weakly on $\tau$, we can neglect the term $\sim c_4$ as $\tau\to0$, resulting in 
\be\label{XXA}
\lambda=\frac{\tau(1+Q)c_2}{c^2_3G^2}=\frac{(1+Q)c_2}{c^2_3}
\frac{\tau(\xi)}{\xi^4 L^{2(D+2)}}.
\ee

For constant $c_2,c_3$ and $Q$ no valley is present in eq. \eqref{XXA}. In this case the solutions correspond to geometrical runaway solutions where $G\to\infty$. Solutions wit $G\to G_0$ therefore require a nontrivial dependence of $Q,c_2,c_3$ on $G$ even in the limit $\tau\to 0$. This will depend on the detailed geometry. In particular, we recall that the field equation for $\xi$ receives an additional contribution from $\partial\tau/\partial\xi$, thus modifying $Q$. For $\tau\to 0$ the geometrical properties leading to a nonzero $G_0$ may become independent of $\tau$, such that $G_0$ approaches a $\tau$-independent constant. The only remaining $\tau$-independence in $\lambda$ is then the overall factor of $\tau$ which drives $\tau\to 0$. We will call the solutions with constant $G_0$ ``anomalous runaway solutions''. They are the analogue of the scaling solutions discussed in the preceding section. The criteria for the existence of a valley solution, namely a minimum of $U_0$ as a function of $G$, are the same for the scaling and the anomalous runaway solutions. In the first case $\tau$ is some nonzero constant, in the second $\tau\to 0$. 

Anomalous runaway solutions can exist both for positive and negative anomalous dimension $A_\tau$. For $A_\tau>0$ one finds asymptotically $\xi\to\infty$ and therefore $L\to 0$ (for fixed $G_0$). In contrast, for $A_\tau<0$ the asymptotic behavior is $\xi\to 0,L\to\infty$. In the latter case the internal space is continuously expanding even in the present cosmological epoch, similar to the geometrical runaway solution. However, this effect is now partially cancelled by the decrease of $\xi$.

Let us next discuss the evolution of the dimensionless combination $\chi L$. With eq. \eqref{48} one has 
\be\label{64}
\chi^2L^2=(c_3G+c_4\tau)=\frac{c_1G+c_2\tau}{c_3G+c_4\tau}\lambda^{-1}.
\ee
For the geometrical runaway solutions with $G\to\infty$ we recover eq. \eqref{55} and therefore the same problem as in the preceding section, namely the issue of a potentially too strong time dependence of the couplings in the standard model. In contrast, for the anomalous runaway solutions this problem is absent. For constant $c_3G_0$ and $\tau\to 0$ one finds $\chi^2L^2\to c_3G_0$. In this case the compactification scale $M_c$ scales proportional to the Planck mass, $M_c\sim \chi$. The dimensionless couplings and mass ratios of the standard model are functions of $G$ and $\tau$. Typically, there may exist a limit $\tau\to 0$, where these couplings only depend on $G$. For $G\to G_0$, and $G_0$ independent of $\tau$, the couplings will go to fixed points. We conclude that anomalous runaway solutions would precisely lead to asymptotic fixed points $\lambda_*=0,~h^2_*\neq 0$ in the sense of the discussion in sect. \ref{dilatationsymmetry}. A residual time variation of the couplings, as discussed in sect. \ref{timevarying}, may arise for solutions where $G_0$ or other geometrical features approach their asymptotic values for $\tau\to 0$ only slowly.

For anomalous runaway solutions the four dimensional anomalous dimension $A$ can be related directly to the higher dimensional anomalous dimension $A_\tau$. With 
\be\label{XXB}
\lambda(\chi)=\bar\lambda\tau(\xi)=\bar\lambda \tau_0\left(\frac{\xi}{\mu}\right)^{-A_\tau}=\tilde\lambda\chi^{-\frac{A_\tau}{2}(D+2)},
\ee
and (almost) constant
\be\label{XXC}
\bar\lambda=\frac{(1+Q)c_2}{c^2_3G^2_0}~,~\tilde\lambda=\bar\lambda \tau_0\mu^{A_\tau}_\tau
G^{\frac{A_\tau}{4}}_0c_3^{\frac{A_\tau}{4}(D+2)},
\ee
we identify
\be\label{XXD}
A=\frac{D+2}{2}A_\tau.
\ee
The solutions where $\chi$ runs asymptotically to infinity, exploring the ultraviolet, correspond to $\xi\to\infty, L\to 0$. In this case, a computation of $A$ involves the ultraviolet behavior of the higher dimensional theory. It seems rather obvious that the four dimensional gluon fluctuations have not much to say about the issue of the cosmological constant, as stated in our answer to the question \eqref{10}. We notice that there are also realistic cosmologies for $A_\tau<0,~A<0$. In this case one has for increasing time $\chi\to 0,~\xi\to 0,~L\to \infty$ such that cosmology explores the infrared limit of the higher dimensional theory. Nevertheless, the dominant contributions for a computation of $A_\tau$ remain in both cases the modes with momenta $p^2\sim\chi^2\sim L^{-2}$. From the point of view of the effective four dimensional theory, these are again the high momentum modes near the Planck mass. The cosmology with $A>0$ and $A<0$ is rather similar \cite{CWAA}, both leading to an exponential potential \eqref{1}. Only the sign in the definition of $\varphi$ changes.

\section{Geometrical runaway for warped branes}
\label{geometricalrunaway}

Warped branes \cite{CWP}, \cite{RSUW} combine the idea that we may live in a type of membrane embedded in a higher dimensional world \cite{RSB}, with a warped geometry \cite{WH}. We propose here that geometrical runaway solutions for warped branes could lead to a realistic cosmology and particle physics, provided the warping is strong and the fermions and gauge bosons of the standard model are concentrated near the brane.

The discussion of the last two sections can be summarized in a simple formula
\ba\label{W1}
\lambda(\chi,L)=\frac{v(\chi,L)}{\chi^2L^2}~,~
v=\frac{c_1G+c_2\tau}{c_3G+c_4\tau},
\ea
with $G=\xi^2L^{D+2}$, and
\ba\label{W2}
c_1&=&\int_y(g^{(0)}_D)^{1/2}\sigma^2\left(\frac{\zeta}{2}\partial^\alpha\xi^{(0)}\partial_\alpha\xi^{(0)}
-\frac12\xi^{(0)2}\tilde R^{(0)}_D\right),\nn\\
c_2&=&\int_y(g^{(0)}_D)^{1/2}\sigma^2(\tilde R^{(0}_D)^{\frac d2},\nn\\
c_3&=&\int_y(g^{(0)}_D)^{1/2}\sigma\xi^{(0)2},\nn\\
c_4&=&-d\int_y(g^{(0)}_D)^{1/2}\sigma(\tilde R^{(0)}_d)^{\frac d2-1}.
\ea
In eq. \eqref{W1} we have replaced $\xi$ by $\chi$ using eq. \eqref{48}, and $\tilde R^{(0)}_D$ is given by eq. \eqref{30}, using $g^{(0)}_{\alpha\beta}$. There are two possible alternatives for obtaining a realistic present value $\lambda(t_0)\approx 10^{-120}$. Either $v$ is of the order one, and a small $\lambda$ obtains by geometrical runaway with $\chi L\to \infty$. Or $v(\chi,L)$ is a tiny quantity. We will explore next scenarios, for which the present value of $L$ can be larger than the millimeter scale and therefore in a range where gravity is well probed.

The large size $L$ of internal space requires a  new length scale $l_b$, and appropriate physics which prevents the internal dimensions to become visible for typical experimental scales $l\gtrsim l_b$. A possible mechanism for ``hiding'' the additional dimensions is strong warping or, more generally, a strong enough concentration of the ``graviton wave function'' in a small region in internal space. Consider the function
\be\label{W3}
\kappa(y)=\sigma(y)g^{1/2}_D(y)\xi^2(y),
\ee
which is the integrand in eq. \eqref{39} for $\mu_\xi$. Assuming that the term $\mu_\tau$ gives at most a contribution of the same size as $\mu_\xi$, we may interprete $\kappa(y)$ as the graviton wave function in internal space. We will assume that the integral $\mu_\xi=\int_y\kappa(y)$ is dominated by a small region of internal space with linear size $l_b$. 

In other words, the graviton wave function is strongly peaked near the brane, which we may locate at $y=0$. We use cartesian coordinates $y^\alpha$ in the vicinity of $y=0$, with $r^2=\delta_{\alpha\beta}y^\alpha y^\beta$, and assume that $\kappa$ depends only on $r$, such that
\be\label{W4}
\mu_\xi\sim\int dr r^{D-1}\kappa(r).
\ee
As an example, we may consider two regions with a qualitatively different $r$-dependence of $\kappa$. For $r\lesssim l_b$ we take
\be\label{W5}
\kappa(r)=\tilde \kappa_1 l_b^{-(D+2-\epsilon_1)}r^{-\epsilon_1},
\ee
where the dimensionless constant $\tilde\kappa_1$ may depend on 
\be\label{W5A}
g_b=\xi^2l^{(D+2)/2}_b,
\ee
with $\xi$ playing the role of an effective four dimensional field as in the preceding section. The central assumption is here that the geometry near the singularity is independent of the overall size $L$ of the space in which the singularity is located. This is similar to known singularities, as black holes. We take $\epsilon_1<D+1$ such that $\mu_\xi$ remains finite. Away from the singularity, for $r\gtrsim r_b$, we assume a similar behavior as eq. \eqref{W5}, but now with constants $\tilde\kappa_2,\epsilon_2$ and $\epsilon_2>D+1$. Then the integral is peaked at $r\approx l_b$ and we obtain
\be\label{W6}
\mu_\xi\sim\tilde\kappa_1 l^{-2}_b.
\ee

As a second assumption, we consider situations where the characteristic scale of $\tilde R_D$ is given by $L,~\tilde R_D=\tilde R^{(0)}_DL^{-2}$, and not by $l_b$. Again this has an analogy for black holes in some cosmological background in four dimensions, where the curvature scalar is given by cosmology, typically $R\sim H^2$, while the geometry near the black hole is governed by a different scale, the Schwarzschild radius. As a consequence, $\mu_\xi$ and $\mu_\tau$ can have vastly different size and we may typically neglect $\mu_\tau$. If $\tilde\kappa_1$ reaches asymptotically a constant for large $t$, one infers
\be\label{W7}
\chi^2\sim l^{-2}_b.
\ee
In terms of the coefficients \eqref{64} this implies
\be\label{W8}
c_3\sim c_4\sim\frac{\tilde\kappa_1}{g_b}\left(\frac{l_b}{L}\right)^D.
\ee
Similarly, we may estimate
\be\label{W9}
\tilde U_\xi\sim \int_y g^{1/2}_D\sigma^2\xi^2\tilde R_D\sim
\int_y\sigma(y)\kappa(y)L^{-2}
\sim\tilde\kappa_1\tilde \sigma_1 l^{-2}_b L^{-2},
\ee
where we assume
\be\label{W10}
\sigma(y)=\tilde\sigma_1 l_b^{\eta_1} r^{-\eta_1}.
\ee
Here $\tilde\sigma_1$ may depend on $g_b$ and we assume $\epsilon_1+\eta_1<D+1$. Dimensional analysis yields $c_1\approx c_2\approx c_3/\tilde\sigma_1$, such that $v$ is indeed of the order one if $\tilde \sigma_1$ is of the order one. A small $\lambda$ can therefore result from geometrical runaway, $\chi L\to\infty$. Furthermore, for $G\to\infty$ the terms $\sim\tau$ become irrelevant. If $c_1\sim G^{-1}$, as suggested by the field equation for $\xi$ (cf. sects. \ref{dilatationsymmetryhigher}, \ref{simplemmodel}), also $v\sim G^{-1}$ can become much smaller than one. In this case a runaway of $\lambda$ towards zero can be realized by a simultaneous increase in $\chi L$ and $G$. 

The new aspects for the geometrical runaway for branes arise from the relation \eqref{W7}, $\chi\sim l^{-1}_b$. Let us imagine that the wave functions of all standard model particles are concentrated near the brane in the sense that only the geometry in the range $0<r\lesssim l_b$ influences the relevant dimensionless couplings and masses. Dimensional analysis implies for particle masses $m$ and dimensionless couplings $h$
\be\label{W11}
m=l^{-1}_b\tilde m(g_b)~,~h=\tilde h (g_b),
\ee
with $\tilde m$ and $\tilde h$ dimensionless functions of the dimensionless combination $g_b$ \eqref{W5A}. For a realistic particle physics it is now sufficient that the asymptotic value for $g_b$ reaches a constant fixed point $g_{b_*}$ (or that $\tilde h,\tilde m,\tilde\kappa_1$ become independent of $g_b$). Then $h$ reaches a fixed point $h_*$ and the running of the dimensionless couplings stops. Similarly, the ratio between particle masses and the Planck mass reaches a constant, $m/\chi\sim\tilde m (g^*_b)$. Also the effective unification mass $M_{GUT}=l^{-1}_b$ is proportional to $\chi$. This constitutes a possible solution to the problem with time varying couplings for the geometrical runaway solutions.

The second problem - the ``large scale problem'' - could also be solved. First of all , if $c_1\sim G^{-1}$ the present value of $L$ is sufficiently small such that internal space is not observable, as discussed in sect. \ref{dilatationsymmetryhigher}. Furthermore, it seems reasonable that the standard model particles are subject to an effective four dimensional gravity for all scales $l>l_b$. The effective unification scale $l^{-1}_b$ can be large, perhaps in the vicinity of the Planck mass, such that solar system or laboratory observations only see standard gravity, even if $L$ turns out large. For the validity of an effective four dimensional gravity it is necessary that the infinitely many four dimensional tensor fields, that result from the higher dimensional metric $\hat g\mn(x,y)$ beyond the four dimensional metric $g^{(4)}\mn(x)$ - the tower of Kaluza-Klein modes - play only a negligible role for the interactions of the standard model particles. 

If the problem of time varying couplings is solved by concentrating the wave functions of the standard model particles and the graviton near the brane, the geometrical runaway becomes a very interesting candidate for a solution of the cosmological constant problem. The effective cosmon potential goes to zero for large $\varphi$, according to eq. \eqref{52} for constant $c_1$, and according to 
\be\label{102A}
U\sim M^4\exp \left(-\frac{2(D+2)}{\kappa}\frac{\varphi}{M}\right)
\ee
for $c_1\sim G^{-1}$, with $\kappa$ determined by the coefficient of the kinetic term for $\ln L$ in the Einstein frame after dimensional reduction. The fact that it goes to zero and not to a constant arises from the runaway of the size of internal space, $L\to \infty$. 

\section{Adjusting internal geometry}
\label{adjustinginternal}
The discussion in the preceeding section relies on the possibility that vastly different scales may characteristic the geometry in the presence of singularities (branes). This phenomenon, which is well-known from the physics of black holes, can arise in a much wider context than discussed in sect. \ref{geometricalrunaway}. The crucial ingredient is the appearance of a scale $\tilde L$ characterizing the curvature $\tilde R_D\sim \tilde L^{-2}$, which is different from the scale $l_b$ which is relevant for particle physics. ``Geometrical adjustment'' describes scenarios where $l_b$ is given by the characteristic size of internal space, which is typically small and assumed to reach a constant value in units of the Planck mass $\chi$. If we keep denoting by $L$ the ``size'' of internal space, we should replace in the formulae of the preceeding section $l_b\to L,~L\to\tilde L$, such that eq. \eqref{W7} becomes now
\be\label{103}
\chi^2\sim L^{-2}.
\ee

The geometrical runaway $L\to\infty$ is now replaced by geometrical adjustment $\tilde L\to\infty$. Again, the effective four dimensional theory will lead to an exponentially vanishing cosmon potential \eqref{102A}, with $\varphi\sim \ln\tilde L$. The contributions of the terms $\sim \tau$ to the effective potential and Planck mass in the reduced four dimensional theory are suppressed by inverse powers of $\tilde L$, since they involve the curvature $\tilde R_D$. The geometrical adjustment describes the approach of a general class of cosmologies to the special solutions discussed in sect. \ref{simplemmodel}. Both for the geometrical runaway and the geometrical adjustment, for which the internal curvature $\tilde R_D$ goes to zero, the effect of the higher order curvature invariants contained in $F$ (eq. \eqref{24}) goes dynamically to zero.

We have seen in sect. \eqref{simplemmodel} that a solution consistent with dimensional reduction for a static internal space has $c_1=0$ and therefore a flat potential for the radion. This flat valley in the potential gets now influenced by terms $c_1\sim \tilde L^{-(D+2)}$. If $L$ and $\xi$ approach asymptotically constant values, the cosmon corresponds to $\ln\tilde L$. Otherwise, the cosmon may correspond to a more complicated combination of $\ln\tilde L,\ln L$ and $\xi$. In all cases, one will find an exponential potential of the type \eqref{1}, with $\lambda_*=0$. We also stress that for the geometrical runaway and the geometrical adjustment the higher dimensional dilatation anomaly is not important. These scenarios can work with or without the running of $\tau$. The four dimensional effective dilatation anomaly appears in the form of an anomalous dimension due to geometry in higher dimensions. 

At this place it seems useful to comment on models that have a more general form of $F$ as given in eq. \eqref{24A}. The geometrical runaway for warped branes could be realized for a rather arbitrary form of $F$. The anomalous runaway and the geometrical adjustment are characterized by an asymptotic vanishing of the contribution of $F$ to the effective cosmon potential. For the anomalous runaway, we have to deal typically with several dimensionless couplings characterizing $F$. They may all simultaneously run towards zero. For example, the ratios of different couplings in $F$ could evolve to zero or nonzero fixed points, such that only the overall coupling $\tau$ matters. For the geometrical adjustment it is necessary that $F$ decays asymptotically with some inverse power of $\tilde L$. It is presumably sufficient that the adjustment is towards higher dimensional Einstein spaces, but a more detailed study seems necessary.

\section{Conclusions}
\label{conclusions}
In the second part of this paper we have discussed three possible ways how higher dimensional physics could lead to an asymptotically vanishing cosmological constant. In the {\em anomalous runaway} scenario, the relaxation of the effective four dimensional cosmon-potential towards zero is induced by a running higher dimensional coupling $\tau$. The anomalous dimension for this coupling in the higher dimensional theory translates into an anomalous dimension for the effective four dimensional cosmological constant $\lambda M^4$. The {\em geometrical runaway for brane cosmologies} realizes the decrease of $\lambda$ by an expansion of the volume $L^D$ of internal space, measured in units of the four dimensional Planck mass. This scenario becomes viable if the particle physics is governed by a brane scale $l_b$ different from $L$, such that the dimensionless couplings and mass ratios approach constants despite the increase of $L$. Finally, in the {\em adjusting geometry} scenario the ratio between the compactification scale $M_c=L^{-1}$ and the Planck mass reaches asymptotically a constant, guaranteeing for stable particle physics properties. The runaway is now realized by a characteristic length scale for the internal curvature $\tilde L$ increasing towards infinity.

All three scenarios illustrate the general discussion in the first part of this paper, which has been done in the four dimensional theory. There are a few central lessons:
\begin{itemize}
\item [(i)]The physics of the cosmological constant is Planck scale physics.
\item [(ii)]Naive estimates of the ``natural size'' of the cosmological constant by a computation of quantum fluctuations of some low energy modes are not meaningful. These computations fail to incorporate the important symmetry properties related to dilatations, or the geometrical features of a higher dimensional world. Furthermore, quantum fluctuations should not be computed in the Einstein frame, but rather in the higher dimensional theory - typically corresponding in the effective four dimensional theory to a type of Jordan frame.
\item[(iii)]Exponential potentials for the cosmon field are natural outcomes of all three higher dimensional scenarios. They reflect the nature of the cosmon as a pseudo-Goldstone boson of dilatation symmetry, broken by anomalies or explicitely. At the same time, they give rise to interesting scaling solutions in cosmology that could explain why the dark energy is at present of a similar size as matter.
\item[(iv)]A crucial problem of higher dimensional cosmologies is the stabilization of the particle physics properties, in the sense that dimensionless couplings and mass ratios should at most very mildly depend on time. The central issue is here the simultaneous runaway of the cosmological constant to zero and the almost constancy of couplings \cite{CWQ1}. Each feature separately can be realized in many models in a simple and natural way. Our three scenarios propose different possibilities how both properties can be realized simultaneously. The essential ingredient is the running of the particle physics couplings towards nonzero fixed points. If this running is fast, no time dependence of the couplings will be observable. For a slow approach to the fixed point, one may detect coupling variations, since the approach continues in the present cosmological epoch. The varying couplings correspond to a ``fifth interaction'' mediated by the cosmon field, that is substantially weaker than gravity. This is observable by apparent violations of the equivalence principle.
\end{itemize}

In  this paper we have restricted ourselves to general aspects of the problem of the cosmological constant. A concrete realization of one of our three higher dimensional scenarios would be most welcome. This requires the computation of the higher dimensional dilatation anomaly and the existence of cosmological solutions with the appropriate geometric properties. The existence of static solutions with warping, that may be the asymptotic state of higher dimensional cosmologies, is well established. The task is now the computation of cosmologies that realize an approach to such an asymptotic state.

\section*{APPENDIX: Brane solutions in higher dimensional Einstein space}
\renewcommand{\theequation}{A.\arabic{equation}}
\setcounter{equation}{0}
In this appendix we discuss special solutions of the field equations for dilatation symmetric higher dimensional actions for constant $\xi$. These are higher dimensional Einstein-spaces. We concentrate on solutions with vanishing four dimensional curvature scalar.

Let us consider a $d$-dimensional Einstein space with metric obeying
\be\label{A1}
\hat R_{\hat \mu\hat\nu}=0.
\ee
Together with $\xi=~const.$ this is a solution of the field equations \eqref{25}, \eqref{26}. We specify the warped metric \eqref{29} by using flat four dimensional space $g^{(4)}\mn=\eta\mn,\sigma=\sigma(z)$ and
\be\label{A2}
g^{(D)}_{\alpha\beta}=L^2
\left(\begin{array}{ccc}
1&,&0\\
0&,&\rho(z)\bar g_{\bar\alpha\bar\beta}(\bar y)
\end{array}\right).
\ee
Here $D-1$ coordinates $\bar y^{\bar\alpha}$ form a homogeneous space, $\bar R_{\bar\alpha\bar\beta}=C\bar g_{\bar\alpha\bar\beta}$. The dimensionless coordinates $\bar y^{\bar\alpha}$ are of the type of angles, such that an overall length scale $L$ can be written as multiplicative constant. Our conventions are such that $C=(D-2)$ for $\bar g_{\bar\alpha\bar\beta}$ parameterizing a unit sphere $S^{D-1}$. We have to determine from the field equation  \eqref{A1} how the dimensionless functions $\sigma(z)$ and $\rho(z)$ depend on the dimensionless coordinate $z$. 

Solutions of this type have been discussed in \cite{RDW} and we follow this work. The field equations (with primes denoting derivatives with respect to $z$)
\ba\label{A3}
\frac{(D-1)C}{\rho}&=&3\frac{\sigma''}{\sigma}+\frac 32(D-1)\frac{\rho'}{\rho}\frac{\sigma'}{\sigma}+(D-1)\frac{\rho''}{\rho}\nn\\
&&+\frac{1}{4}(D-1)(D-4)\left(\frac{\rho'}{\rho}\right)^2,\nn\\
\frac{(D-3)C}{\rho}&=&(D-2)\frac{\rho''}{\rho}+\frac14(D-2)(D-5)\left(\frac{\rho'}{\rho}\right)^2\nn\\
&&+2(D-2)\frac{\rho'}{\rho}\frac{\sigma'}{\sigma}+4\frac{\sigma''}{\sigma}-\left(\frac{\sigma'}{\sigma}
\right)^2,
\nn\\
\frac{(D-1)C}{\rho}&=&2(D-1)\frac{\rho'}{\rho}\frac{\sigma'}{\sigma}+3\left(\frac{\sigma'}{\sigma}\right)^2\nn\\
&&+\frac14(D-1)(D-2)\left(\frac{\rho'}{\rho}\right)^2,
\ea
contain only two independent equations for the two functions $\sigma(z)$ and $\rho(z)$. For the special case $D=2,~C=0$ these are the field equations used for the first warped solutions in ref. \cite{WH}, but in absence of a higher dimensional cosmological constant. The eqs. \eqref{A3} can be combined into a second order differential equation for $Y$
\be\label{A4}
Y''+\frac{4}{D-1}Y^{-(D-1)}Y'+\frac{(D+3)}{(D-1)^2}Y^{-2D+3}=0,
\ee
with
\be\label{A5}
\frac{\sigma'}{\sigma}=-Y^{-(D-1)}~,~
\frac{\rho'}{\rho}=-\frac{2}{D-1}\left(\frac{\sigma'}{\sigma}+\frac{\sigma''}{\sigma'}\right).
\ee

For $D>2$ the general solution of eq. \eqref{A4} corresponds to the damped motion of a particle in an effective potential
\be\label{A5A}
V_{eff}=-\frac{D+3}{2(D-2)(D-1)^2}Y^{-2(D-2)},
\ee
and we note the special role of $D=2$, where the potential becomes logarithmic, $V_{eff}=5\ln Y$. One may obtain the different types of solutions by fixing at some $z_{in}>0$ the values $Y(z_{in})$ and $Y'(z_{in})$, corresponding to two free integration constants. We consider first $Y'(z_{in})>0$. For $z<z_{in}$, nothing can prevent the solution from reaching the origin at $Y=0$ for some value $z_0$. We may choose $z_0=0$, thus fixing one of the integration constants. As $z$ increases, $Y$ first increases. The motion either reaches a turning point, and $Y$ subsequently decreases until it reaches zero, with $Y(z\to \bar z)\to 0$. Alternatively, the increase of $z$ may continue for $z\to\infty$, with $Y'(z\to\infty)=A>0$. The second integration constant corresponds to $\bar z$ or $A$. As a boundary case, one has a turning point at infinity, $\bar z\to\infty$ or $A\to 0$. 

The solutions with
\be\label{A6}
\lim_{z\to 0}Y=0
\ee
are singular for $z\to 0$, since $\sigma'/\sigma$ diverges. They correspond to a ``brane'' located at $z=0$. Close to the singularity one finds
\be\label{A7}
Y=\left(\frac{z}{\eta_+}\right)^{\frac{1}{D-1}}~,~\sigma=\bar\sigma z^{-\eta_+}~,~\rho=\bar\rho z^{\gamma_+},
\ee
with
\ba\label{A8}
\eta_+&=&\frac{1}{D+3}\left(\sqrt{(D-1)(D+2)}-2\right)~,~\nn\\
\gamma_+&=&\frac{2+4\eta_+}{D-1}.
\ea
For $D>2$ one has $\eta_+>0$, with $\eta_+(D\to 2)=0$. We note $\rho^{\frac{D-1}{2}}\sigma\sim z^{1+\eta_+}$ such that the integral $c_3$ \eqref{W2} is well behaved for $z\to 0$ and the Planck mass is finite. The solutions with a turning point have a second singularity at $\bar z$, with
\be\label{A8A}
Y=\left(\frac{z-\bar z}{\eta_-}\right)^{\frac{1}{D-1}}~,~\sigma=\bar\sigma_-(\bar z-z)^{-\eta_-}~,~
\rho=\bar\rho_-(\bar z-z)^{\gamma_-}
\ee
and
\ba\label{A8B}
\eta_-&=&-\frac{1}{D+3}(\sqrt{(D-1)(D+2)}+2)~,~\nn\\
\gamma_-&=&\frac{2+4\eta_-}{D-1}.
\ea
From $\rho^{\frac{D-1}{2}}\sigma\sim(\bar z-z)^{1+\eta_-}$ one concludes that the integral $c_3$ is finite in this case.

On the other hand, the solution with $Y'>0$ for $z\to\infty$ reads, for $D>2$
\ba\label{A9}
Y&=&Az+B~,~\sigma=\sigma_\infty\exp\left\{
\frac{(Az+B)^{2-D}}{A(D-2)}\right\},\nn\\
\rho&=&\left(z+\frac BA\right)^2\exp 
\left\{-\frac{4(A z+B)^{2-D}}{(D-2)(D-1)A},\right\},\nn\\
\ea
were we have chosen a normalization of standard ``polar coordinates'' for flat space, i.e. $\rho(z\to\infty)=z^2$. For $D\geq3$ a reasonable approximation for large $z$ is simply 
\be\label{A10}
\sigma=1~,~\rho=z^2,
\ee
where we have chosen a particular normalization of $\sigma$. For $\bar g_{\alpha\beta}$ the metric of a unit sphere $S^{D-1}$, we recognize that this solution describes flat space ${\mathbbm R}^d$, with a brane located at $z=0$ deforming the geometry locally. For $\sigma_\infty>0$ the $z$-integral for the computation of the coefficient $c_3$ diverges - no effective four dimensional description of gravity is possible. The boundary case between the solutions \eqref{A8A} and \eqref{A9} is given by the solution \eqref{A7}, \eqref{A8}, which is actually an exact special solution valid for all $z$. Even though $\sigma(z\to\infty)\to 0$, no effective four dimensional theory is valid since $c_3$ is not given by a convergent integral for $z\to \infty$. 

The solutions with $Y'(z_{in})<0$ always approach a singularity at $\bar z>z_{in}$, which is of the type \eqref{A8A}, \eqref{A8B}. Again the solution \eqref{A8A} is an exact ``dividing solution'', which now extends to $z\to-\infty$. For the dividing solution we can compute the critical value $Y'(z_{in})=Y'_c<0$. For $Y'(z_{in})>Y'_c$ we have a turning point for $z<z_{in}$ and find a second brane described by eqs. \eqref{A7}, \eqref{A8}. This is the same type of solutions with two singularities as discussed for $Y'(z_{(in)}>0$. On the other hand, for $Y'(z_{in})$ smaller than $Y'_c$ the solutions extends now to $z\to-\infty$, with $\sigma\to\sigma_\infty>0$ and $Y\to\infty$ according to 
\be\label{A14}
Y\to -(Az+B).
\ee
Only the solutions with two branes lead to a valid four dimensional description with finite $c_3$. Of course, the role of the two branes can be interchanged by a reflection in the $z$-coordinate.

\end{document}